\documentclass[journal,transmag]{IEEEtran}
\usepackage{caption}
\usepackage[switch]{lineno}
\usepackage[cmex10]{amsmath}
\usepackage{amsfonts}
\usepackage{algorithm, algorithmic}
\usepackage{array}
\usepackage{etoolbox,xstring,mfirstuc,textcase}
\usepackage{mdwmath}
\usepackage{mdwtab}
\usepackage{eqparbox}
\usepackage{mathrsfs}
\usepackage{tabularx}
\usepackage{amsthm}
\usepackage{amsmath,bm}
\usepackage{booktabs}
\usepackage{multirow}
\usepackage{diagbox}
\usepackage{txfonts}
\usepackage{bm}
\usepackage{makecell}
\usepackage{subfigure}
\usepackage{graphicx}
\usepackage{enumitem}
\usepackage{cite}
\usepackage{ltxtable}
\usepackage{amsmath}
\usepackage{amssymb}
\usepackage{mathrsfs}
\usepackage{setspace}
\usepackage{multirow}
\usepackage{subfigure}
\usepackage{hyperref} 
\usepackage{acronym}
\usepackage{easyReview}
\usepackage[normalem]{ulem} 

\usepackage[utf8]{inputenc}
\usepackage[english]{babel}

\definecolor{mypink1}{rgb}{0.858, 0.188, 0.478}
\definecolor{green(pigment)}{rgb}{0.0, 0.65, 0.31}
\definecolor{blue(munsell)}{rgb}{0.0, 0.5, 0.69}
\definecolor{spirodiscoball}{rgb}{0.06, 0.75, 0.99}
\definecolor{cinnamon}{rgb}{0.82, 0.41, 0.12}
\definecolor{mypink1}{rgb}{0.858, 0.188, 0.478}
\definecolor{ao}{rgb}{0.0, 0.0, 1.0}
\newcommand{\myComment}[1]{{\color{black}{#1}}}
\newcommand{\Hilbertx}{{\cal H}_\mathbf{x}}
\newcommand{\Hilberty}{{\cal H}_\mathbf{y}}
\newcommand{\kernely}{{k}_\mathbf{y}}
\newcommand{\kernelx}{{k}_\mathbf{x}}
\newcommand{\featurey}{\phi_\mathbf{y}}
\newcommand{\featurex}{\phi_\mathbf{x}}
\newcommand{\pdf}{p}
\newcommand{\Gramx}{K_{\mathbf{x}\mathbf{x}}}
\newcommand{\GramxT}{K_{\tilde{\mathbf{x}}\mathbf{x}}}
\newcommand{\GramxTT}{K_{\tilde{\mathbf{x}}\tilde{\mathbf{x}}}}
\newcommand{\Gramy}{G_{\mathbf{y}\mathbf{y}}}

\def\R{\mathbb{R}}

\usepackage{dirtytalk}
\usepackage{multicol}
\usepackage[none]{hyphenat}

\title{Adaptive Kernel Kalman Filter}

\author{Mengwei Sun, \IEEEmembership{Member, IEEE}, 
Mike E. Davies, \IEEEmembership{Fellow, IEEE}, Ian K. Proudler,\\
James R. Hopgood, \IEEEmembership{Senior Member, IEEE}% and Third C. Author, Jr., \IEEEmembership{Member, IEEE}
\thanks{M. W. Sun, M. E. Davies and J. R. Hopgood are with Institute 
of Digital Communications, University of Edinburgh, Edinburgh, EH9 3FG, U.K. E-mail: (msun; mike.davies; james.hopgood)@ed.ac.uk.

I. Proudler is with  the Centre for Signal \& Image Processing (CeSIP), Department of Electronic \& Electrical Engineering, University of Strathclyde, Glasgow, G1 1XW, U.K. E-mail: 
ian.proudler@strath.ac.uk.

This work was supported by the Engineering and Physical Sciences Research Council (EPSRC) Grant number EP/S000631/1; and the MOD University Defence Research Collaboration (UDRC) in Signal Processing.

For the purpose of open access, the author has applied a Creative Commons Attribution (CC BY) licence to any Author Accepted Manuscript version arising from this submission.
}
}

\begin{document}
%\linenumbers
\maketitle
%\tableofcontents

\acrodef{KME}{kernel mean embedding}
\acrodef{DSSM}{dynamic state-space model}
\acrodef{pdf}{probability density function}
\acrodef{EKF}{extended Kalman filter}
\acrodef{AKKF}{adaptive kernel Kalman filter}
\acrodef{KKR}{kernel Kalman rule}
\acrodef{GPF}{Gaussian particle filter}
\acrodef{RKHS}{reproducing kernel Hilbert space}
\acrodef{UKF}{unscented Kalman filter}
\acrodef{PF}{particle filter}
\acrodef{GPF}{Gaussian particle filter}
\acrodef{BOT}{bearing-only tracking}
\acrodef{KF}{Kalman filter}
\acrodef{KBR}{kernel Bayes' rule}
\acrodef{UNGM}{univariate nonstationary growth model}
\acrodef{MSE}{mean square error}
\acrodef{CV}{constant velocity}
\acrodef{LMSE}{logarithmic mean square error}
\acrodef{AWGN}{additive white Gaussian noise}
\acrodef{MC}{Monte Carlo}
\acrodef{CT}{coordinated turn}

\begin{abstract}
Sequential Bayesian filters in  non-linear dynamic systems require the recursive estimation of the predictive and posterior \myComment{\acfp{pdf}}. 
This paper introduces a Bayesian filter called the \ac{AKKF}.
\myComment{The \ac{AKKF} approximates the arbitrary predictive and posterior \acp{pdf} of hidden states using the \acp{KME} in \acp{RKHS}.}
In parallel with the \acp{KME}, some particles in the data space are used to capture the properties of the dynamic system model. Specifically, particles are generated and updated in the data space. \myComment{Moreover, the corresponding kernel weight means vector and covariance matrix associated with the particles' kernel feature mappings are predicted and updated in the \acp{RKHS} based on the \ac{KKR}.} Simulation results are presented to confirm the improved performance of  our approach with significantly reduced \myComment{numbers of particles} by comparing with the \ac{UKF}, \ac{PF}, and \ac{GPF}. {For example}, compared with the \ac{GPF}, 
the  \ac{AKKF}  provides  around $50\%$ \ac{LMSE} tracking  performance improvement  in the \ac{BOT} system when using $50$ particles.
\end{abstract} 

\begin{IEEEkeywords}
\textit{Adaptive kernel Kalman filter, Non-linear dynamic systems, Sequential Bayesian filters, Kernel mean embedding,  Kernel Kalman rule.}
\end{IEEEkeywords}
\acresetall

\section{Introduction}
Many problems in the fields of science,  including statistical signal processing, target tracking, and satellite navigation, require parameter estimation in non-linear dynamic systems. In order to make \myComment{inferences} about a discrete-time  dynamic  system, \myComment{a \ac{DSSM}  is  required},  including  a process model describing the evolution of the hidden states with time, as shown in \eqref{eqn:DSM1}, and  a  measurement  model  relating  the  observations  to  the states,  as shown in \eqref{eqn:DSM2}{{;}}
\begin{align}
\label{eqn:DSM1}
    \mathbf{x}_n &= \mathit{f}\left(\mathbf{x}_{n-1}, \mathbf{u}_{n}\right),\\
   \label{eqn:DSM2}    
    \mathbf{y}_n &= \mathit{h}\left(\mathbf{x}_{n}, \mathbf{v}_n\right).
\end{align}
Here, $\mathbf{x}_{n}$ represents the hidden state at the $n$-th time slot, $n = 1,\dots,N$, 
{$\mathbf{y}_n$} is the corresponding observation. The process and measurement noise  are represented as $\mathbf{u}_{n}$  and $\mathbf{v}_{n}${,} respectively. \myComment{The process function is  $f: \R^{n_x} \times \R^{n_u} \rightarrow \R^{n_x}$, where $n_x$ and  $n_u$ are the dimensions of the state and process noise vectors, respectively. The measurement function is ${h}: \R^{n_x} \times \R^{n_v} \rightarrow \R^{n_y}$, where $n_y$ and $n_v$ are the dimensions of the observation and measurement noise vectors, respectively.}  In this paper, we introduce a sequential Bayesian filter called the \acf{AKKF} that provides a new view of the approach to state estimation in non-linear dynamic systems.

\subsection{State of the Art --- Non-linear Filters }
\myComment{From a Bayesian perspective on} dynamic state estimation, estimation problems are solved by constructing the posterior \acf{pdf} of hidden states based on all available information, including \acp{DSSM} and received measurements. For problems where a real-time estimate is required after a measurement is received, sequential Bayesian filters are commonly used by recursively computing the posterior \acp{pdf} of the hidden states \cite{9116757,529783,882463}. \myComment{Historically, the main focus of sequential Bayesian filters has been on model-based systems with explicit formulations of \acp{DSSM} \cite{9116757,Gordon1995BayesianSE}. More recently, data-driven Bayesian filters have been proposed where \acp{DSSM} are unknown or partially known, but training data examples are provided \cite{7389490,Kanagawa2014MonteCF, 9272174}}. In both scenarios, the filters are broken down into essentially two stages, \myComment{i.e.,} prediction and update. The predictive \ac{pdf} of the states is calculated in the prediction stage, which is then modified {to become} the posterior \ac{pdf} based on the latest received observation in the update stage  \cite{978374}.

For the model-based filters, 
the predictive  \ac{pdf} of $\mathbf{x}_n$ is obtained in the prediction stage using the process model via the Chapman–Kolmogorov equation \cite{6507656} as
\begin{align}
    \pdf\left(\mathbf{x}_n|\mathbf{y}_{1:n-1}\right) = \int \pdf \left(\mathbf{x}_n|\mathbf{x}_{n-1}\right)\pdf(\mathbf{x}_{n-1}|\mathbf{y}_{1:n-1})d\mathbf{x}_{n-1},
\end{align}
where $\pdf \left(\mathbf{x}_n|\mathbf{x}_{n-1}\right)$ is the state-transition \ac{pdf} defined by the process  model \eqref{eqn:DSM1}{,} $\pdf(\mathbf{x}_{n-1}|\mathbf{y}_{1:n-1})$ is the posterior \ac{pdf}  at time $n-1$.
Then, the updated posterior \ac{pdf}  is  proportional to the product of the measurement likelihood and the predictive \ac{pdf} as \cite{978374}
\begin{align}
  \pdf(\mathbf{x}_n|\mathbf{y}_{1:n}) = \frac{\pdf(\mathbf{y}_{n}|\mathbf{x}_{n})
  \pdf(\mathbf{x}_{n}|\mathbf{y}_{1:n-1})}{\pdf(\mathbf{y}_{n}|\mathbf{y}_{1:n-1})}, 
\end{align}
where $\pdf(\mathbf{y}_{n}|\mathbf{x}_{n})$ is the likelihood function defined by the measurement model \eqref{eqn:DSM2} and the denominator is a normalization term given by:
\begin{align}
{ \pdf(\mathbf{y}_{n}|\mathbf{y}_{1:n-1})=\int \pdf(\mathbf{y}_{n}|\mathbf{x}_{n})\pdf(\mathbf{x}_{n}|\mathbf{y}_{1:n-1})d{\mathbf{x}}_{n}}.
\end{align}

The \ac{KF} \cite{9116757} provides the optimal Bayesian solution for linear DSSMs when the predictive and posterior \acp{pdf} are Gaussian. For state estimation in non-linear systems, the \ac{EKF} \cite{EKF} is \myComment{a popular method} to approximate a recursive maximum likelihood (ML) estimate of the hidden state. 
\myComment{The EKF uses the first derivatives to approximate the process and measurement functions by linear equations.}
However, this can cause poor approximation performance when the model is highly non-linear or when the posterior distributions are multi-modal.
The \ac{UKF} was proposed in \cite{10.111712.280797} as an alternative to the \ac{EKF}. The \ac{UKF} uses a weighted set of deterministic  particles  (so-called  sigma  points) in the state space to approximate the state distribution rather than the \ac{DSSM}. The sigma  points are propagated through the non-linear system to capture the predictive/posterior mean and covariance that is accurate to the third-order of the Taylor expansion \cite{529783,882463}. The underlying philosophy is that
the approximation of a Gaussian distribution with a finite number of parameters is \myComment{more accessible} than the approximation of an arbitrary non-linear function/transformation \cite{Julier96ageneral}. 
Compared with the \ac{EKF}, the \ac{UKF} can significantly improve the accuracy of the approximations. However, divergence can still occur in some non-linear problems as  the  state \acp{pdf}  are  essentially  approximated  as Gaussian \cite{Ito2000GaussianFF}\cite{1232326}.
  
A more general solution to the non-linear Bayesian filter can be found in the sequential \acf{MC} filter, or the \ac{PF} \cite{978374, article}. Similarly to the \ac{UKF}, the \ac{PF} represents the hidden state distributions through a weighted set of points or particles. However, unlike the \ac{UKF}, the particles of the \ac{PF} are chosen and updated stochastically. Specifically, the popular bootstrap \ac{PF}  uses  random  particles  with  associated  weights, \myComment{i.e.,} $\{\mathbf{x}_n^{\{i\}},w_n^{\{i\}}\}_{i=1}^M$, to characterize the posterior \ac{pdf} as% \cite{978374}
\begin{align}
  p\left(\mathbf{x}_n|\mathbf{y}_{1:n}\right) \approx \sum_{i=1}^M w_n^{\{i\}} \delta\left(\mathbf{x}_n - \mathbf{x}_n^{\{i\}}\right){,}
\end{align}
where $\delta(\cdot)$ is the Dirac delta function{,}
$M$ represents the number of particles used at a given time. The key steps of the bootstrap \ac{PF} include:  1) Draw particles from the importance density; 2) Update the particles' weights based on the latest received observation;  3) Particle resampling \cite{978374}. \myComment{Resampling is a necessary step to reduce degeneracy. However, it induces an increase in complexity and is hard to parallelize \cite{861864,7079001}.
In \cite{861864, 1634806, 978374, 6302204,6419797}, various authors proposed specific variants of the bootstrap \ac{PF} to avoid resampling by approximating the hidden state distribution at each time index with a Gaussian.}
 These variants include the \ac{GPF} \cite{861864}, the quasi-Monte Carlo filter \cite{1634806},
the square-root quadrature Kalman filter \cite{978374}, the multiple quadrature Kalman filter \cite{6302204}, and the Gauss–Hermite filter \cite{6419797}. %{\color{red} Drawbacks?}

Different from the model-based approaches above, 
\myComment{several recent works \cite{6530747,Fukumizu13} developed nonparametric data-driven Bayesian filters based on the \ac{KBR}.
These papers represented the \acp{pdf} as a weighted sum of feature vectors in \acp{RKHS} owing  to  the virtue  of  \acp{KME}.}
In \cite{6530747}, the \ac{KBR} was used to derive a kernel Bayesian filter where the evolution of hidden states and the measurement model are unknown and need to be inferred from prior training data.  Subsequent works \cite{7389490} and  \cite{Kanagawa2014MonteCF} have proposed \ac{KBR}-based filters for when the measurement model is only provided through examples of state-observation pairs while the process model is known. \myComment{These papers combine  the parametric \ac{MC} sampling and the nonparametric measurement model learning.}
Specifically, particles are propagated using the process model.  Then, the  posterior \ac{pdf} is approximated by the \ac{KBR} \cite{6530747}. A variant of the \ac{KBR} called the  \ac{KKR} was formulated in \cite{Gebhardt} to overcome some of the instabilities that can be observed in using the \ac{KBR}. These \ac{KME}-based methods can effectively deal with problems that involve unknown measurement models or strong non-linear structures \cite{Kanagawa2014MonteCF}. However, the feature space for the kernel embeddings remains restricted to the training data set. Therefore, the performance of these data-driven filters relies heavily on the sufficient similarity between the training data and the test data, a problem common to all such methods \cite{Fukumizu13}.

\subsection{Novelties and Contributions}

Inspired by the \ac{KBR} \cite{6530747} and \ac{KKR} \cite{Gebhardt}, we introduce a full model-based Bayesian filter called the \acf{AKKF}. \myComment{The work presented in this paper has been built on the
preliminary work in \cite{9541455, Fusion2021} but presents detailed theoretical explanations, a wider set of applications, and computational complexity analysis. }
The main contributions of this paper can be summarized as follows:
\begin{enumerate}
\item We explore the potential of using \acp{KME} within model-based filters. \myComment{The proposed \ac{AKKF} provides a means of utilizing nonparametric data-driven filters within a model-driven framework without the need for any training data or an offline training process. Specifically, 
the AKKF adaptively draws new particles based on \acp{DSSM} to capture the diversity of the non-linearities. 
The kernel embeddings of updated state particles can be seen as providing an adaptive change of basis for the high-dimensional \acp{RKHS}.
Then, the predictive and posterior \acp{pdf} are embedded into the \acp{RKHS} and updated linearly.}
%{and the inferred measurement and transition operators are used to calculate the update rules.}
\item We show that, like the \ac{GPF}, the proposed filter can avoid the resampling step {found} in most \acp{PF}.  However, unlike the \ac{GPF}, it is not constrained to approximate the hidden state \acp{pdf} as Gaussian. %In passing, we also highlight a missing link between the \ac{UKF} sigma point method and the kernel conditional embedding operator.
\item  \myComment{The proposed \ac{AKKF} is tested on three different non-linear systems and compared with the \ac{UKF}, an oracle \ac{PF}, and the \ac{GPF} to demonstrate its efficacy. The tracking
performance and computational complexity comparisons show that the \ac{AKKF} achieves higher accuracy while requiring fewer particles.} For example, compared with the \ac{GPF}, the \acf{LMSE} tracking performance is improved around by $50\%$ in the \ac{BOT} system \myComment{with the target moving following the linear \ac{CV} model}, given $50$ particles.
\end{enumerate}

\begin{figure}[]
	\begin{center}
		\subfigure[] {\label{fig:04a}
			\includegraphics[width=7.5
			cm,height = 3.9cm]{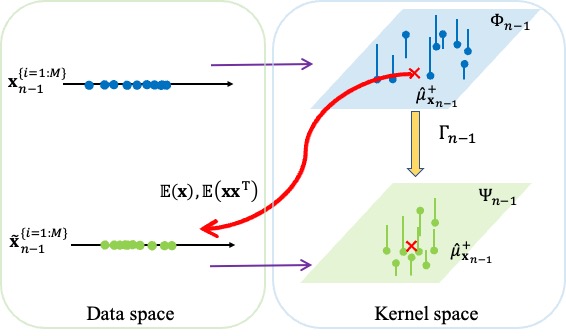}}	
		\subfigure[] {\label{fig:04b}
			\includegraphics[width=7.5
			cm,height = 3.5cm]{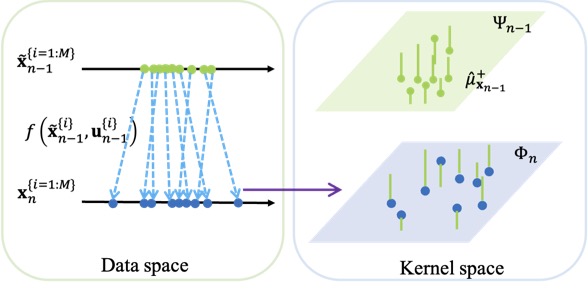}}
				\subfigure[] {\label{fig:041c}
			\includegraphics[width=7.5
			cm,height = 3.8cm]{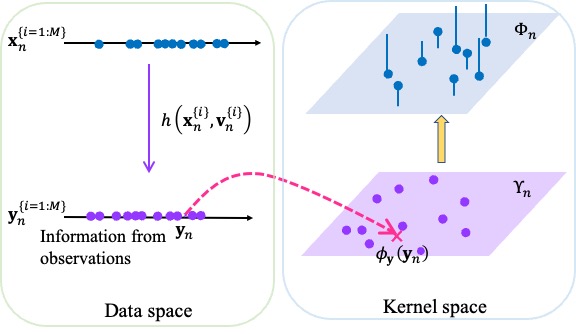}}
		\caption{\label{fig:AKKFmodel} \myComment{ One iteration of the AKKF. Here, $M$ represents the number of particles. (a) Proposal step: embedding the posterior distribution at time $n-1$.
  %$\hat{\mu}_{\mathbf{x}_{n}-1}^{+} = \Phi_{n-1}\mathbf{w}_{n-1}^{+}$. 
  Draw the proposal particles $\uline{\mathbf{x}}_{n-1}^{\{i=1:M\}}$ in the data space according to the importance distribution  accessed  from $\hat{\mu}_{\mathbf{x}_{n-1}}^+$. Then, the proposal kernel weight mean  vector and covariance matrix are updated in the kernel space. (b) Prediction step:  the particles are propagated through process function in the data space. Then the KME of the predictive distribution is approximated as $\pdf(\mathbf{x}_n|\mathbf{x}_{1:n-1},\mathbf{y}_{1:n-1}) \rightarrow \hat{\mu}_{\mathbf{x}_{n}}^{-}$.}  (c) Update step: the information from the new observation is used to update the kernel weight mean  vector and covariance matrix. The \ac{KME} of the  posterior \ac{pdf} $\pdf(\mathbf{x}_n|\mathbf{x}_{1:n-1},\mathbf{y}_{1:n}) \rightarrow \hat{\mu}_{\mathbf{x}_{n}}^{+}$ is calculated based on the observation information.}
	\end{center}
\end{figure}

Compared with the available filters, there are several significant differences and novelties of the proposed \ac{AKKF}:
\begin{enumerate}
  \item \myComment{The state vector’s mean and covariance (in data space) are extracted from the \ac{KME} of the posterior \ac{pdf} for drawing proposal particles, as shown in the proposal step of Fig.~\ref{fig:04a}.} Unlike most of the kernel-based methods where the focus is on characteristic kernels \cite{6530747,Gebhardt}, we also consider simple quadratic and quartic kernels that provide direct access to the mean and covariance of the hidden state. %\myComment{that we can avoid the need for the use of characteristic kernels by simply approximating the DSSM as linear in the feature space.}
    \item  \myComment{The proposal particles can then be precisely propagated through the non-linearity and used to calculate empirical transition operators in the \ac{RKHS} on the fly, as shown in the prediction step of Fig.~\ref{fig:04b}.} Then, those particles' feature mappings with associated kernel weights are used in the kernel feature space to approximate the KME of the posterior \ac{pdf}, see the update step in Fig.~\ref{fig:041c}.  
  Unlike the bootstrap PF and its extensions, the particle weights can take arbitrary values and are not constrained to be non-negative or to sum to one.
    \item  By embedding \acp{pdf} into an \ac{RKHS}, the use of the kernel function allows 
    the statistical inference in non-linear systems to be solved using linear algebra operations. Here the weighted kernel mean vector and weighted kernel covariance matrix are predicted and updated using the \ac{KKR}, \myComment{i.e.,} the \ac{KKR} is used to realize an unbiased update of the \ac{KME} \cite{Gebhardt}.
\end{enumerate}

The rest of the paper \myComment{is} set out as follows. Section II reviews the KME  \cite{6530747}  and the KKR \cite{Gebhardt}. Section III is devoted to the theoretical derivation of the proposed AKKF. In Section IV, we use \myComment{three} typical examples to present the performance results of the \ac{AKKF} for non-linear problems and finally draw conclusions in Section V.

\textit{List of Abbreviations:}

\begin{tabular}{l  l}
    AKKF &Adaptive kernel Kalman filter\\
    BOT &Bearings-only tracking\\
    CV &Constant-velocity\\
    DSSM &Dynamic state-space model\\
    EKF &Extended Kalman filter\\
    GPF &Gaussian particle filter\\
    KBR &Kernel Bayesian rule\\
    KF &Kalman filter\\
    KME &Kernel mean embedding\\
    KKR &Kernel Kalman rule\\
    LMSE & Logarithmic mean square error\\
    ML &Maximum-likelihood\\
    MSE &Mean square error\\
    PF &Particle filter\\
    RKHS &Reproducing kernel Hilbert space\\
    UKF &Unscented Kalman filter\\
    UNGM &Univariate nonstationary growth model \\
\end{tabular}

\section{Preliminaries}
This section briefly reviews the framework of the \ac{KME}, empirical \ac{KME}, and data-driven \ac{KKR}. See \cite{6530747} and \cite{Gebhardt} for details.

\subsection{Kernel Mean Embedding}

A random variable is denoted as $X$ in the data space $\mathcal{X}$ with a \ac{pdf} $\pdf(\mathbf{x})$.  An instance of $X$ is denoted as $\mathbf{x}$.
A \acf{RKHS} denoted as $\Hilbertx$ on the data space $\mathcal{X}$ with a kernel function $\kernelx(\mathbf{x},\mathbf{x}')$ is defined as a Hilbert space of functions {$f(\cdot)$} with the inner product $\langle \cdot,\cdot\rangle_{\Hilbertx}$ that satisfies the following important properties:
\begin{itemize}
    \item The feature mapping of $\mathbf{x}$: $ \featurex(\mathbf{x}) = \kernelx(\mathbf{x} , \cdot) \in \Hilbertx$ for all $\mathbf{x} \in \mathcal{X}$.
    \item Reproducing property: $ f(\mathbf{x}) = 	\langle f,\kernelx(\mathbf{x},·)\rangle_{\Hilbertx}$ for all $f \in \Hilbertx $ and $\mathbf{x} \in \mathcal{X}$.
\end{itemize}
\myComment{The above definitions are also applied to the predecessor of current state ${X}$, i.e., $\uline{X}$, and the observation variable, i.e., $Y$, that sit in two \acp{RKHS}. See Table I for a summary.} The kernel function is the inner product of two feature mappings, \myComment{i.e.,}  $\langle \featurex(\mathbf{x}),\featurex(\mathbf{x}')\rangle _{\Hilbertx}= \kernelx(\mathbf{x}, \mathbf{x}')$.
Table II  gives some typical kernel functions assuming a scalar $\mathbf{x}$. 
\myComment{This paper investigates both infinite-dimension and finite-dimension feature spaces in a common framework.} 
\begin{table}[!t]
\small
    \centering
        \caption{The meaning of notations.}
    \begin{tabular} {p{2.2cm}|p{1.6cm}|p{1.6cm}| p{1.6cm}}
    \hline\hline
Description &  {Current state} &  {predecessor state} & Observation\\ \hline 
Random variable & $X$ & $\underbar{X}$  & $Y$   \\ \hline 
Domain  & $\mathcal{X}$ & $\mathcal{X}$ & $\mathcal{Y}$\\ \hline 
Specific variable  &  $\mathbf{x}$ & $\uline{\mathbf{x}}$ &$\mathbf{y}$\\ \hline 
Kernel    & $\kernelx(\mathbf{x},\mathbf{x}')$ & $\kernelx(\uline{\mathbf{x}},\uline{\mathbf{x}}')$   & $\kernely(\mathbf{y},\mathbf{y}')$  \\ \hline
Kernel matrix & $\Gramx$
 &  $K_{\uline{\mathbf{x}}\uline{\mathbf{x}}}$ & $\Gramy$  \\ \hline 
 Feature mapping & $\featurex(\mathbf{x})$
 &  $\featurex(\uline{\mathbf{x}})$ & $\featurey(\mathbf{y})$   \\ \hline 
 Feature matrix & $\Phi$ & $\Psi$ &  $\Upsilon$    \\ \hline 
 \ac{RKHS} & $\Hilbertx$&  $\Hilbertx$ &  $\Hilberty$  \\ \hline\hline
    \end{tabular}
    \label{tab:my_label}
\end{table}

The \ac{KME} approach represents a \ac{pdf} $\pdf(\mathbf{x})$ by an element in the RKHS as
\begin{align}
    \mu_X := \mathbb{E}_X \left[ \featurex(X)\right] = 
    \int_{\mathcal{X}} \featurex(\mathbf{x}) \pdf(\mathbf{x})d \mathbf{x}.
    \label{eqn:1kernelem1}
\end{align}
%The mean embedding $\mu_X $ has the property that the expectation of any RKHS function $f$ can be evaluated as an inner product in $\mathcal{F}$, $\langle \mu_X, f\rangle _{\mathcal{F}} := \mathbb{E}_X \left[f(X)\right] \forall f \in  \mathcal{F}$. 

The joint \ac{pdf} of two or more variables, \myComment{e.g.,} $\pdf(\mathbf{x},\mathbf{y})$, can be embedded  into a tensor product feature space $\Hilbertx \otimes \Hilberty$ as a (uncentered) covariance operator, {\footnote{While some results have been formulated with centered kernels, \myComment{e.g.,} \cite{Fukumizu04}, equivalent derivations can be made for the uncentered covariance operator.}} \myComment{i.e.,}
\begin{equation}
    \begin{aligned}
    \mathcal{C}_{XY} :&= \mathbb{E}_{XY} \left[ \featurex(X)\otimes
    \featurey(Y)\right]\\
    &= \int_{\mathcal{X}\times \mathcal{Y}} \featurex(\mathbf{x}) \otimes \featurey(\mathbf{y})\pdf(\mathbf{x},\mathbf{y})  d \mathbf{x} d\mathbf{y}.
    \label{eqn:1kernelem2}
\end{aligned}
\end{equation}
The tensor product features satisfy 
$\langle \featurex(\mathbf{x})\otimes
    \featurey(\mathbf{y}), \featurex(\mathbf{x}')\otimes
    \featurey(\mathbf{y}') \rangle_{\Hilbertx \otimes \Hilberty} = \kernelx(\mathbf{x},\mathbf{x}') \kernely(\mathbf{y},\mathbf{y}')$.
    
 Similar to (\ref{eqn:1kernelem1}), the \ac{KME} of a conditional \ac{pdf} $\pdf(X|\mathbf{y})$ can be defined as
\begin{align}
    \mu_{X|\mathbf{y}} := \mathbb{E}_{X|\mathbf{y}} \left[ \featurex(X)\right] = \int_{\mathcal{X}} \featurex(\mathbf{x}) \pdf(\mathbf{x}|\mathbf{y}) d\mathbf{x}.
     \label{eqn:1kernelem3}
\end{align}
The difference between the KME  {$\mu_{X|\mathbf{y}}$} and $\mu_{X} $ is that $\mu_{X} $ is a single element in the RKHS, while  {$\mu_{X|\mathbf{y}}$} is a family of points, each indexed by fixing $Y$ to a particular value $\mathbf{y}$.
A conditional operator $\mathcal{C}_{X|Y}$ is defined under certain conditions \cite{Fukumizu13} as the linear operator,  which takes the feature mapping of a fixed value $\mathbf{y}$ as the input and outputs the corresponding conditional KME; 
\begin{align}
\label{eq:eqKME10}
    \mu_{X|\mathbf{y}} = \mathcal{C}_{X|Y} \featurey(\mathbf{y}) = \mathcal{C}_{XY}\mathcal{C}^{-1}_{YY} \featurey(\mathbf{y}).
\end{align}
In practice, 
\myComment{it is difficult to make valid statistical inferences about the regression parameters with an ill-conditioned covariance operator.}
Therefore, the inversion of $\mathcal{C}_{YY}$ is \myComment{generally} replaced by the regularized inverse, \myComment{i.e.,}
$\left(\mathcal{C}_{YY} + {\kappa} I\right)^{-1}$, where $\kappa$ is a regularization parameter to ensure that the inverse is well-defined, \myComment{and $I$ is the identity operator matrix.} When the conditions defined in \cite{Fukumizu13} are not \myComment{precisely} met, \eqref{eq:eqKME10} can still be interpreted as a linear (in the feature space) minimum mean squared error estimate for $\mu_{X|y}$.
\begin{table}[!t]
\small
     \begin{center}
        \caption{ {Typical kernel functions.}}
    \begin{tabular}{p{1.7cm}|p{3cm}|p{2.4cm}}
    \hline
    \hline 
Kernel function & $k({\mathbf{x}}, {\mathbf{x}'})$ & Dimension of feature mapping   \\ \hline 
 \hline 
 Linear & $\langle {\mathbf{x}}, {\mathbf{x}'} \rangle$ &  Finite \\ 
 \hline 
Quadratic  & $ \left(\langle {\mathbf{x}}, {\mathbf{x}'} \rangle + c\right)^2$
 &  Finite\\ \hline 
Quartic   & $\left(\langle {\mathbf{x}}, {\mathbf{x}'} \rangle + c\right)^4$&
Finite\\ \hline 
Gaussian   & $\exp\left( -\frac{1}{\sigma^2} \|({\mathbf{x}}-{\mathbf{x}'})\|^2\right)$   & Infinite
\\ \hline
\hline 
    \end{tabular}
    \label{tab:my_label}
    \end{center}
\end{table}

\subsection{\myComment{Empirical Estimation of KME}}

As it is rare to \myComment{access the actual} underlying \acp{pdf} mentioned above, we can alternatively estimate the \acp{KME} using a finite number of samples drawn from
the corresponding \acp{pdf}.

The empirical \ac{KME} of $\pdf(\mathbf{x})$ 
in \eqref{eqn:1kernelem1} is approximated as the average of the samples' feature mappings, i.e., $\mathcal{D}_{\featurex(X)} = \{ \featurex(\mathbf{x}^{\{1\}}), \dots, \featurex(\mathbf{x}^{\{M\}})\}$, the samples $\mathcal{D}_X = \{ \mathbf{x}^{\{1\}}, \dots, \mathbf{x}^{\{M\}}\}$ are drawn i.i.d. from $\pdf(\mathbf{x})$, and $M$ represents the number of samples; 
\begin{align}
    \hat{\mu}_X = \frac{1}{M}\sum_{i=1}^M\featurex(\mathbf{x}^{\{i\}}).
\end{align}

The empirical KME of the covariance operator $\mathcal{C}_{XY}$ in \eqref{eqn:1kernelem2}  inherits the injectivity of $\mathcal{C}_{XY}$ and is approximated as
\begin{align}
     \hat{\mathcal{C}}_{XY} = \frac{1}{M}\sum_{i=1}^M \featurex(\mathbf{x}^{\{i\}})\otimes \featurey(\mathbf{y}^{\{i\}})=\frac{1}{M}\Phi\Upsilon^{\rm{T}},
\end{align}
where the  $M$ sample pairs $\mathcal{D}_{XY} = \{ (\mathbf{x}^{\{1\}},\mathbf{y}^{\{1\}}), \dots, (\mathbf{x}^{\{M\}},\mathbf{y}^{\{M\}})\}$  are drawn i.i.d. from $\pdf(\mathbf{x},\mathbf{y})$ with the feature mappings 
$\Phi := \left[\featurex(\mathbf{x}^{\{1\}}),\dots,\featurex(\mathbf{x}^{\{M\}})\right]$
and $\Upsilon := \left[\featurey(\mathbf{y}^{\{1\}}),\dots,\featurey(\mathbf{y}^{\{M\}})\right]$.\footnote{For infinite dimensional feature spaces these operators are infinite-dimensional.  {However, a practical implementation} is still possible working in the data space and using the kernel trick. For finite-dimensional feature spaces, empirical calculations can be implemented in either the feature space or using the kernel trick in the data space. }
%\ikpc{I think it would help the reader if we stated what the dimensions of $\Phi$ and $\Upsilon $ are.}

The KME of the conditional distribution $\pdf(\mathbf{x}|\mathbf{y})$ is theoretically calculated as \eqref{eqn:1kernelem3} {or \eqref{eq:eqKME10}}.
When $\pdf(\mathbf{x}|\mathbf{y})$ is unknown but i.i.d. samples $\mathcal{D}_{XY}$ drawn from $\pdf(\mathbf{x},\mathbf{y})$ are available, the estimation of the empirical conditional operator  $\hat{\mathcal{C}}_{X|Y}$ is regarded as a linear regression in the RKHS \cite{9541455, Fusion2021}. 
%see the illustrated blue line in Fig.~\ref{fig:figKME} \cite{9541455, Fusion2021}. 
And $\hat{\mathcal{C}}_{X|Y}$  is calculated by
  \begin{equation}
  \label{eq:eqAKKF_condcov}
  \begin{split}
 \hat{\mathcal{C}}_{X|Y} = \hat{\mathcal{C}}_{XY} \left(\hat{\mathcal{C}}_{YY} + \kappa  I\right)^{-1}
 = \Phi \left(\Gramy + \kappa I\right)^{-1} \Upsilon^{\rm{T}}.
 \end{split}
 \end{equation}
 Here, $\Gramy = \Upsilon^{\rm{T}} \Upsilon $ is the Gram matrix for the  samples from the observed variable $Y$. 
\myComment{Then, the empirical \ac{KME} of the conditional distribution is calculated through the following linear algebra as}
\begin{gather}
    \hat{\mu}_{X|\mathbf{y}} = \hat{\mathcal{C}}_{X|Y}
    \phi(\mathbf{y})=
     \Phi\left(\Gramy + \kappa I\right)^{-1} \Upsilon^{\rm{T}}\featurey(\mathbf{y}) {\equiv} \Phi\mathbf{w}.
\end{gather}
%\ikpc{As the last step defines $\mathbf{w}$ I think it would be better  to have $\hat{\mu}_{X|\mathbf{y}} = $ \dots $\equiv \Phi\mathbf{w}$}
Here, $\mathbf{y} \in \mathcal{Y}$ is the input test variable. Note \myComment{that} the empirical \ac{KME} of the conditional \ac{pdf} $\hat{\mu}_{X|\mathbf{y}}$ is a weighted sum of the feature mappings of the training samples. The weight vector includes $M$ non-uniform weights, \myComment{i.e.,} ${\mathbf{w}} = \left[ w^{\{1\}}, \dots, w^{\{M\}} \right]^{\rm{T}}$, and is calculated as
\begin{gather}
     \label{eq:eqKME_W}
   \mathbf{w} =  \left(\Gramy + \kappa I\right)^{-1} G_{:,\mathbf{y}}, 
   \end{gather}
where the vector of kernel functions $G_{:,\mathbf{y}} = \left[ \kernely(\mathbf{y}^{\{1\}}, \mathbf{y}), \dots, \kernely(\mathbf{y}^{\{M\}}, \mathbf{y}) \right]^{\rm{T}}$. {From \eqref{eq:eqKME_W}, we can {see} that the kernel weight vector is the solution to a set of linear equations in the feature space, \myComment{and unlike \ac{PF} methods there are} no non-negativity or normalization constraints.}
%\begin{figure}[!t]
%	\centering
%	\includegraphics [width=5cm,height=7.5cm]{KME.jpg}
%	\caption{ KME of the conditional distribution $\pdf(X|\mathbf{y})$ is embedded as a point in an \ac{RKHS} as $\mu_{X|\mathbf{y}} = \int_{\mathcal{X}} \phi(\mathbf{x})\pdf(\mathbf{x}|\mathbf{y}) d\mathbf{x}$. Given the training data sampled from $\pdf(X,Y)$, the
%	empirical KME of $\pdf(X|\mathbf{y})$ is approximated as a linear operation in RKHS, \myComment{i.e.,}  $\hat{\mu}_{X|\mathbf{y}}=\hat{C}_{X|Y}\phi(\mathbf{y})=\Phi \mathbf{w}$. Legend: $\cdot$ samples, $\times$: empirical \ac{KME}, $*$: \ac{KME}. }
%	\label{fig:figKME}
%\end{figure}

\subsection{Kernel Kalman Rule}

Based on the \ac{KME} of conditional \acp{pdf}, non-linear estimations can be mapped into kernel feature spaces, \myComment{i.e., \acp{RKHS},} and solved using linear operations.
It has been proposed that the conditional \ac{KME} operator in \eqref{eq:eqKME10} can then be used to derive a \ac{KBR} under certain conditions \cite{6530747,Fukumizu13}. However, these conditions are \myComment{very restrictive} and often fail, making the formulation difficult to interpret theoretically and quite unstable practically.  Recently an alternative,  the \acf{KKR} has been proposed  exploiting the optimal linear interpretation (in the kernel feature space) of the conditional \ac{KME} estimate that enjoys better stability \cite{Gebhardt}.

\begin{figure}[!t]
	\centering
	\includegraphics [width=8.5cm,height=7cm]{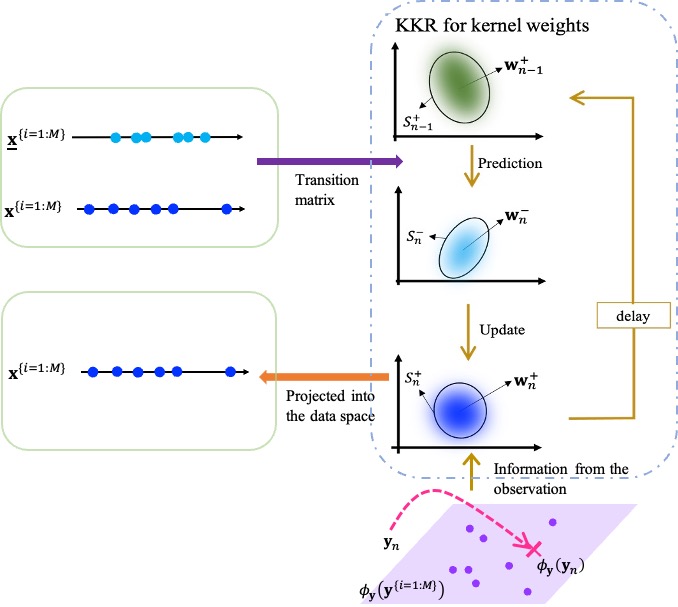}
	\caption{One iteration of the \ac{KKR}. Here, $\uline{\mathbf{x}}^{\{i=1:M\}}$, $\mathbf{x}^{\{i=1:M\}}$, and $\mathbf{y}^{\{i=1:M\}}$ denotes the deterministic training predecessor hidden states, current hidden states, and observations, respectively. The kernel weight mean vector and covariance matrix are predicted and updated as a KF in kernel spaces. The estimations of hidden states are found by projecting the kernel weights into the training data space. 1. Prediction step: The predictive kernel weight vector and matrix are updated based on the transition matrix $T$.  2. Update step: The posterior kernel weight vector and matrix are updated according to the information of the new observation $\mathbf{y}_n$.}
	\label{fig:figKKR}
\end{figure}

The empirical \ac{KKR} is formulated by executing a \ac{KF} in the kernel feature space. \myComment{An illustration of the \ac{KKR} is shown in  Fig.~\ref{fig:figKKR}.
Specifically, the transition matrix and the transition residual \myComment{are} calculated based on the training data set that is assumed to include  $M$ sample {triples} $\mathcal{D}_{\uline{X}XY} = \{ (\uline{\mathbf{x}}^{\{1\}},\mathbf{x}^{\{1\}},\mathbf{y}^{\{1\}}), \dots, (\uline{\mathbf{x}}^{\{M\}},\mathbf{x}^{\{M\}},\mathbf{y}^{\{M\}})\}$ \cite{Gebhardt}. Here, $\uline{\mathbf{x}}^{\{i\}}$ denotes the predecessor state of ${\mathbf{x}}^{\{i\}}, i =1,\dots,M$, and $\mathbf{y}^{\{i\}}$ is the \myComment{corresponding} observation of ${\mathbf{x}}^{\{i\}}$. The feature mappings of the training data are represented as $\uline{\Phi} := \left[\featurex(\uline{\mathbf{x}}^{\{1\}})\dots,\featurex(\uline{\mathbf{x}}^{\{M\}})\right]$,  
${\Phi} := \left[\featurex({\mathbf{x}}^{\{1\}})\dots,\featurex({\mathbf{x}}^{\{M\}})\right]$ and ${\Upsilon} := \left[\featurey(\mathbf{y}^{\{1\}})\dots,\featurey(\mathbf{y}^{\{M\}})\right]$, respectively. 
The corresponding kernel weight mean vector and its covariance matrix are 
calculated following the prediction and update steps in the weight space. }

In the prediction step, the kernel weight vector and covariance matrix from time $n-1$ to  time $n$ are predicted in the weight space as
\begin{gather}
\mathbf{w}^{-}_{n} =  T  \mathbf{w}^{+}_{n-1},\\
S^{-}_{n} =  T  S^{+}_{n-1} T^{\rm{T}}+V.
\end{gather}
Here, the kernel transition matrix $T$ is calculated based on the training predecessor states and training states data as
$T = \left( K_{\uline{\mathbf{x}}\uline{\mathbf{x}}}+ {{\lambda}_{\uline{K}}} I \right)^{-1}K_{\uline{\mathbf{x}}\mathbf{x}}$,  
{and $V$ represents the transition residual \cite{KKR}}. ${K}_{\uline{\mathbf{x}}\mathbf{x}} = \uline{\Phi}^{\rm{T}}\Phi$ is the transition Gram matrix, and ${K}_{\uline{\mathbf{x}}\uline{\mathbf{x}}}= \uline{\Phi}^{\rm{T}}\uline{\Phi}$ is the Gram matrix of the predecessor states, {${\lambda}_{\uline{K}}$ is the regularization parameter to stabilize the inverse of ${K}_{\uline{\mathbf{x}}\uline{\mathbf{x}}}$}. The predictive \ac{KME} and covariance operator estimates are {then} calculated as $\mu_{\mathbf{x}_n}^{-}=\Phi\mathbf{w}^{-}_{n}$ and $\mathcal{C}_{\mathbf{x}_n\mathbf{x}_n}^{-}=\Phi S^{-}_{n} \Phi^{\rm{T}}$, respectively.

{Next}, the innovation update is executed based the kernel Kalman gain $Q_{n}$ calculation in the update step, \myComment{i.e.,}
\begin{gather}
    \mathbf{w}^{+}_{n} = \mathbf{w}^{-}_{n} +  Q_{n} \left(
    {G_{:,\mathbf{y}_n}}-G_{\mathbf{y}\mathbf{y}}\mathbf{w}^{-}_{n} \right),\\
    S^{+}_{n} = S^{-}_{n} -  Q_{n}G_{\mathbf{y}\mathbf{y}}S^{-}_{n},\\
        Q_{n} = S^{-}_{n} \left(
    G_{\mathbf{y}\mathbf{y}}S^{-}_{n} + \kappa I \right)^{-1},
\end{gather}
where the Gram matrix of the training observations is $G_{\mathbf{y}\mathbf{y}} = \Upsilon^{\rm{T}}\Upsilon$. The test observation at time $n$ is $\mathbf{y}_n$,  the kernel function vector between the  training observations and the test observation is $G_{:,\mathbf{y}_n} = \left[ \kernely(\mathbf{y}^{\{1\}}, \mathbf{y}_n), \dots, \kernely(\mathbf{y}^{\{M\}}, \mathbf{y}_n) \right]^{\rm{T}}$. \myComment{The updated \ac{KME} and covariance operator estimates are then calculated as $\mu_{\mathbf{x}_n}^{+}=\Phi\mathbf{w}^{+}_{n}$ and $\mathcal{C}_{\mathbf{x}_n\mathbf{x}_n}^{+}=\Phi S^{+}_{n} \Phi^{\rm{T}}$, respectively.}
\myComment{If the \ac{KME} contains linear functions, e.g., when quadratic or quartics kernels are used, we can directly calculate the mean and covariance of the hidden states in the data space as marginal quantities of the estimated \ac{KME}. 
Even when this is not possible, e.g., as with Gaussian kernels, a good approximation can be obtained by projecting the estimated \ac{KME} into the data space as \eqref{eq:eqKKRPro1} and \eqref{eq:eqKKRPro2} \cite{KKR} where it is implicitly assumed that functions in kernel feature spaces can reasonably approximate the linear and quadratic functions;}
\begin{gather}
\label{eq:eqKKRPro1}
\hat{\mathbf{x}}_n = X_{\mathcal{D}} \mathbf{w}^{+}_{n},\\
\label{eq:eqKKRPro2}
\hat{\Sigma}_n = X_{\mathcal{D}} S^{+}_{n} X_{\mathcal{D}}^{\rm{T}},
\end{gather}
where $X_{\mathcal{D}} = \left[ \mathbf{x}^{\{1\}},\dots,\mathbf{x}^{\{M\}}\right]$ is the set of the current training states.

\myComment{
The kernel-based filters learn the probabilistic transition and observation dynamics as linear functions on embeddings of the belief state in high-dimensional \acp{RKHS} from training data. Note that existing filters based on the KME or the KKR are entirely data-driven, requiring the training data to provide sufficient statistics of the dynamic systems and, therefore, of use when the \ac{DSSM} is unavailable. The tracking applications of the \ac{KKR} so far include table tennis balls track, human motion activity estimation, and pendulum track \cite{Gebhardt}. These applications all have the weakness that the high-dimensional \acp{RKHS} are limited with the training data, which requires high similarities between the test data and the training data. 
However, the entirely data-driven filter's tracking performance is vulnerable and will fail catastrophically when the target moves out of the training space.
This is particularly a problem in the case for the real-time tracking applications that we focus on here. To the best of our knowledge, other investigations have not considered the issue of incorporating a \ac{DSSM} into the \ac{RKHS} setting.  

Unlike the \ac{KKR}, this paper proposes a Bayesian filter called the \acf{AKKF} that provides a mechanism for applying the data-driven kernel method to model-based systems. Specifically, there is no need for any training data or an offline training process of the \ac{AKKF}. 
The \ac{AKKF} adaptively draws new particles whose weighted features match the current \ac{KME} estimate. These particles can then be precisely propagated through the non-linearity 
and used to calculate empirical transition operators in the \ac{RKHS} on the fly. The embeddings of updated state particles can be seen as providing an adaptive change of basis for the high-dimensional RKHSs, making the non-linear function approximation more accurate and flexible. Therefore, the \ac{AKKF} has higher efficiency and broader applications. }

\section{Adaptive Kernel Kalman Filter}

The proposed \ac{AKKF} {aims} to take all the benefits  of the \ac{KME} and  \ac{KKR}, and adapt them to work in the model-based setup. \myComment{I.e.,} the presented \ac{AKKF} is a method incorporating  a \ac{DSSM} into \acp{RKHS}.
In a similar manner to \myComment{the} selection and  propagation of sigma points in the \ac{UKF}, the \ac{AKKF} adaptively updates particles whose weighted features are matched to the \ac{KME} estimate of the current state.  \myComment{Note that the AKKF chooses particles propagated through the non-linear system randomly,  which is different from the \ac{UKF}. Further, the weights of the proposed \ac{AKKF}, unlike \acp{PF}, do not need to be normalized or non-negative and are updated through simple linear regression.} 

In the proposed AKKF, the empirical KME of the hidden state's posterior \ac{pdf}
requires a set of generated particles' feature mappings and the corresponding kernel weights. 
\myComment{Fig.~\ref{fig:AKKFmodel} shows one iteration of the proposed \ac{AKKF} executed in both the data and kernel feature spaces. Specifically, particles are updated and propagated in the data space based on parametric \acp{DSSM} to capture the diversity of the non-linearities. The corresponding kernel weight mean vector and covariance matrix are predicted and updated 
{by matching (or approximating in a least squares manner in the feature space) with the state \ac{KME}.}
The following presents three main steps of the proposed \ac{AKKF}. }

%\begin{figure*}[!h]
%	\centering
%	\includegraphics [width=16cm,height=11cm]{Image/AKKF.png}
%	\caption{{ \color{blue}One iteration of the AKKF. The particles are updated in data space. The particles and proposal particles are propagated and updated in data space based on the DSSM, as shown in (a1)--(a4).
%	The empirical KME is approximated by the feature mappings of states with the associated Kernel weights in Kernel feature space, as shown in (b1)--(b4). The Kernel weight vector with matrix are predicted and updated as a Kernel Kalman filter process in Kernel space, as shown in (c1)--(c4). }}
%	\label{fig:figuoverlay-implement}
%\end{figure*}

\subsection{Embedding the Posterior Distribution at Time $n-1$}

%The posterior kernel weight mean vector and covariance matrix at the preceding time slot $n-1$ are represented as $\mathbf{w}^{+}_{n-1}$ and ${S}^{+}_{n-1}$, respectively, as shown in Fig. 4(a). And the corresponding feature mappings of particles with weights in the kernel feature space are shown in Fig. 3(b1) as black points and green arrows, respectively. }
%In the data space, the particles at preceding time slot $n-1$ are represented as $\{ x_{n-1}^{\{1\}},\dots,x_{n-1}^{\{M\}}\}$, as shown in  Fig. 2(a1). The corresponding feature mappings with weights in the Kernel feature spaceare shown in Fig. 2(b1) as black points and green arrows, respectively.

Given the posterior \ac{KME} estimate at time $n-1$, i.e., $\hat{\mu}^{+}_{\mathbf{x}_{n-1}}$, we wish to draw new particles that better represent the probability mass of the associated posterior \ac{pdf}. \myComment{The posterior \ac{KME} estimate $\hat{\mu}^{+}_{\mathbf{x}_{n-1}}$
comprises weighted feature mappings of the particles, 
for which we use blue points to represent in Fig.~\ref{fig:04a}.}  
While there are sophisticated iterative methods, such as herding \cite{Chen2010SuperSamplesFK}, that can sample from the posterior distribution. \myComment{We advocate a much simpler technique in the spirit of importance sampling. Given that we can extract estimates for the mean and covariance of the state \ac{pdf} in data space, we can draw particles in the high probability region of the \ac{pdf} by sampling from a Gaussian distribution with matched mean and covariance. These particles can then be used to generate a new approximation of the \ac{KME} of the \ac{pdf} through appropriate reweighting.}

Specifically, the particles and the corresponding kernel feature mappings at time slot $n-1$ are represented as $\mathbf{x}^{\{i=1:M\}}_{n-1}$ and $\featurex(\mathbf{x}^{\{i=1:M\}}_{n-1})$, respectively.
%as the blue points shown in Fig.~\ref{fig:04a}.
And the empirical \ac{KME} and the 
covariance operator of  $p(\mathbf{x}_{n-1}\mid \mathbf{x}_{0:n-1}, \mathbf{y}_{1:n-1}) $ were calculated as
\begin{align}
\label{eq:eqAKKF1_1}
    \hat{\mu}^{+}_{\mathbf{x}_{n-1}} %=\mathbb{E}\left[\featurex(X_{n-1})\right]
    &=\Phi_{n-1}\mathbf{w}^{+}_{n-1},\\
    \label{eq:eqAKKF1_2}
    \hat{\mathcal{C}}_{\mathbf{x}_{n},\mathbf{x}_{n}} &= %\mathbb{E}\left[\featurex(X_{n-1})\otimes\featurex(X_{n-1})  \right] = 
    %\begin{split}
   \Phi_nS_n^{+}\Phi_n^{\rm{T}},
      %&= \Phi_{n-1} \mathrm{diag}(\mathbf{w}^{+}_{n-1})\Phi_{n-1}^{\rm{T}} - \hat{\mu}_{n-1} \hat{\mu}_{n-1}^{\rm{T}},\\
      %&=\Phi_{n-1} \mathrm{diag}(\mathbf{w}^{+}_{n-1})\Phi_{n-1}^{\rm{T}} -  \left[\Phi_{n-1} \mathbf{w}^{+}_{n-1}\right]  \left[\Phi_{n-1} \mathbf{w}^{+}_{n-1}\right] ^{\rm{T}},\\
      %&=\Phi_{n-1} \mathrm{diag}(\mathbf{w}^{+}_{n-1})\Phi_{n-1}^{\rm{T}} -  \Phi_{n-1} \mathbf{w}^{+}_{n-1} \mathbf{w}^{+}_{n}^{\rm{T}} \Phi_n^{\rm{T}},\\
      %&=\Phi_n\left[ \mathrm{diag}(\mathbf{w}^{+}_{n}) -\mathbf{w}^{+}_{n} \mathbf{w}^{+}_{n}^{\rm{T}} \right]\Phi_n^{\rm{T}},\\
      %&=\Phi_nS_n^{+}\Phi_n^{\rm{T}}.
      % \end{split}
\end{align}
where $\Phi_{n-1} = \left[\featurex(\mathbf{x}_{n-1}^{\{1\}}),\dots,\featurex(\mathbf{x}_{n-1}^{\{M\}})\right]$.  % , represents the feature mappings of the posterior particles. %$\mathbf{w}^{+}_{n-1}$  and ${S}^{+}_{n-1}$ are the Kernel weight vector and matrix respectively, as shown in Fig. 2(c1). 
Then, the state mean and covariance (in data space) of $\mathbf{x}_{n-1}$,  \myComment{i.e.,} $\mathbb{E}(\mathbf{x}_{n-1})$ and {$\mathrm{Cov}(\mathbf{x}_{n-1})$}, are extracted from $\hat{\mu}^{+}_{\mathbf{x}_{n-1}}$ and
returned to the data space, as shown by the red arrow in Fig.~\ref{fig:04a}.

\myComment{The state vector’s mean and covariance  are extracted in two different ways: 1) A suitable kernel choice, i.e., quadratic and quartic kernels, can directly give the state vector’s mean and covariance if the associated \ac{RKHS} contains linear functions. 
For example, suppose  $\mathbf{x}_{n-1} = \left[{x}_{n-1,1},\dots,{x}_{n-1,d} \right]^{\rm{T}}$ is a $d$-dimension vector, with the utilization of quadratic  kernel,  the empirical \ac{KME} $\hat{\mu}_{\mathbf{x}_{n-1}}^{+}$ is represented as \eqref{eq:eqKME_2} which contains all features of degree zero, degree one, and degree two terms;
\begin{align}
\label{eq:eqKME_2}
&\hat{\mu}_{\mathbf{x}_{n-1}}^{+} =  \left[\mathrm{vec}\left(\mathbb{E}{\left(\mathbf{x}_{n-1}\mathbf{x}_{n-1}^{\rm{T}}\right)}\right)^{\rm{T}}, \left(\mathbb{E}{\left(\mathbf{x}_{n-1}\right)}\right)^{\rm{T}}, c\right]^{\rm{T}}.%_{d^2+d+1}
\end{align}
Here, $c  \geq 0$ is a free parameter trading off the influence from higher-order and lower-order terms of the polynomial \cite{C6415770}. The utilization of quartic kernel can further provide all features of degree zero to degree four terms;
2) Otherwise, such as linear or Gaussian kernels, the state vector’s mean and covariance 
can be approximated using \eqref{eq:eqKKRPro1} and \eqref{eq:eqKKRPro2}.}
Then, the proposal particles, shown as \myComment{green points} in  Fig.~\ref{fig:04a} can be randomly sampled from the following normal distribution as
\begin{align}
\label{eq:eqproposaldraw}
    \tilde{\mathbf{x}}_{n-1}^{\{i=1:M\}} &\sim \mathcal{N}\left( \mathbb{E}\left(\mathbf{x}_{n-1}\right),  \mathrm{Cov}\left(\mathbf{x}_{n-1}\right)  \right),\\ \mathrm{Cov}\left(\mathbf{x}_{n-1}\right)&= \mathbb{E}\left(\mathbf{x}_{n-1}\mathbf{x}_{n-1}^{\rm{T}}\right)-\mathbb{E}\left(\mathbf{x}_{n-1}\right)\mathbb{E}\left(\mathbf{x}_{n-1}\right)^{\rm{T}}.
\end{align}
\myComment{For convenience, we draw the proposal particles from a Gaussian distribution, although other distributions with matched statistics could also conceivably be used. The proposal particles should therefore capture the location of the significant probability mass of the posterior \ac{pdf}. 
In order to use these particles to approximate the KME of the posterior pdf, we need to calculate new kernel weights for them, i.e,  $\tilde{\mathbf{w}}^{+}_{n-1}$. Note that this is not equivalent to approximating the posterior pdf by a Gaussian. Instead, it can be thought of as an adaptive change of basis within the feature space which can be achieved through a simple linear mapping that we describe next. Let the proposal particles' feature mappings be represented as}
 $\Psi_{n-1} = \left[\featurex(\tilde{\mathbf{x}}_{n-1}^{\{1\}}),\dots,\featurex(\tilde{\mathbf{x}}_{n-1}^{\{M\}})\right]$,
with the associated weight vector $ \tilde{\mathbf{w}}^{+}_{n-1}$ and matrix $\tilde{S}^{+}_{n-1}$. 
Then, the \ac{KME} and covariance operator in \eqref{eq:eqAKKF1_1} and \eqref{eq:eqAKKF1_2} are rewritten as
\begin{align}
    %\Psi_{n-1} &= \left[{\phi}(\tilde{x}_{n-1}^{1}),\dots,\phi(\tilde{x}_{n-1}^{M})
    %\right],\\
    \label{eq:eqAKKF_proposal_weightmean}
    \hat{\mu}^{+}_{\mathbf{x}_{n-1}}&= \Psi_{n-1} 
    \tilde{\mathbf{w}}^{+}_{n-1},\\
    \label{eq:eqAKKF_proposal_weightcov}
       \hat{\mathcal{C}}^{+}_{\mathbf{x}_{n-1}\mathbf{x}_{n-1}}&= \Psi_{n-1} 
     \tilde{{S}}^{+}_{n-1} \Psi_{n-1} ^{\mathrm{T}}.
\end{align}
{The formulas for} the proposal kernel weight vector $ \tilde{\mathbf{w}}^{+}_{n-1} $ and matrix $\tilde{S}^{+}_{n-1}$  are \eqref{eq:eqAKKF_proposal_1} and \eqref{eq:eqAKKF_proposal_2}, respectively.
\begin{equation}
\label{eq:eqAKKF_proposal_1}
\begin{aligned}
   &\Psi_{n-1}
    \tilde{\mathbf{w}}^{+}_{n-1}  =\Phi_{n-1}\mathbf{w}^{+}_{n-1}\\
    \Rightarrow &\Psi_{n-1}^{\mathrm{T}} \Psi_{n-1}
    \tilde{\mathbf{w}}^{+}_{n-1} = \Psi_{n-1}^{\mathrm{T}}\Phi_{n-1}\mathbf{w}^{+}_{n-1}\\
    \Rightarrow 
    &\tilde{\mathbf{w}}^{+}_{n-1} = 
    \left(\Psi_{n-1}^{\mathrm{T}}
    \Psi_{n-1} \right)^{-1}
    \Psi_{n-1}^{\mathrm{T}}\Phi_{n-1}\mathbf{w}^{+}_{n-1}\\
    \Rightarrow 
    &\tilde{\mathbf{w}}^{+}_{n-1} = \left(\GramxTT+ {\lambda_{\tilde{K}}} I \right)^{-1} \GramxT \mathbf{w}^{+}_{n-1}=\Gamma_{n-1} \mathbf{w}^{+}_{n-1},
\end{aligned}
\end{equation}
\begin{equation}
\label{eq:eqAKKF_proposal_2}
\begin{aligned}
&\Psi_{n-1}
\tilde{{S}}^{+}_{n-1}\Psi_{n-1}^{\mathrm{T}}
=\Phi_{n-1}{S}^{+}_{n-1}\Phi_{n-1}^{\mathrm{T}}\\
\Rightarrow 
&\Psi_{n-1}^{\mathrm{T}} \Psi_{n-1}
 \tilde{{S}}^{+}_{n-1}\Psi_{n-1}^{\mathrm{T}}\Psi_{n-1}
=\Psi_{n-1}^{\mathrm{T}} \Phi_{n-1}{S}^{+}_{n-1}\Phi_{n-1}^{\mathrm{T}}\Psi_{n-1}\\
\Rightarrow 
&\tilde{S}^{+}_{n-1} =  \left[\left(\GramxTT+{\lambda_{\tilde{K}}} I \right)^{-1} \GramxT\right]
{S}^{+}_{n-1} \left[\left(\GramxTT+{\lambda_{\tilde{K}}} I \right)^{-1}\GramxT\right]^{\mathrm{T}}\\
\Rightarrow 
&\tilde{S}^{+}_{n-1} = \Gamma_{n-1} {S}^{+}_{n-1} \Gamma_{n-1} ^{\mathrm{T}}.
\end{aligned}
\end{equation}
Here, $\Gamma_{n-1} $ 
represents the change of {sample representation} from $\Phi_{n-1}$ to $\Psi_{n-1}$
and is calculated as $\Gamma_{n-1} = \left(\GramxTT+ {\lambda_{\tilde{K}}} I \right)^{-1} \GramxT$. The matrix $\GramxTT = \Psi^{\rm{T}}_{n-1} \Psi_{n-1}$  represents the Gram matrix of the proposal particles at time $n-1$.  The matrix 
$\GramxT= \Psi^{\rm{T}}_{n-1} \Phi_{n-1}$ represents the Gram matrix between the {old particles $\mathbf{x}^{\{i=1:M\}}_{n-1}$ and the proposal particles $\tilde{\mathbf{x}}^{\{i=1:M\}}_{n-1}$} at time $n-1$.   The regularization parameter  $\lambda_{\tilde{K}}$ is used to stabilize the inverse of $\GramxTT$.
Note that for small feature spaces, \myComment{i.e.,} $\Phi$ is full  rank, and $\mathrm{Dim} \left[\Phi\right] < M$,   \eqref{eq:eqAKKF_proposal_weightmean} 
and 
\eqref{eq:eqAKKF_proposal_weightcov} are exact. However, to deal with ill-conditioning or where the feature space is larger than the number of samples, \myComment{e.g.,} when it is infinite, {using the weight vector} and  covariance matrix from \eqref{eq:eqAKKF_proposal_1} and \eqref{eq:eqAKKF_proposal_2} make \eqref{eq:eqAKKF_proposal_weightmean} and \eqref{eq:eqAKKF_proposal_weightcov} approximate.

\subsection{Prediction from Time $n-1$ to Time $n$}

In this step, the proposal particles generated in the previous step are propagated through the process model to estimate the  transition operator ${\mathcal{C}}_{\mathbf{x}_{n}|\mathbf{x}_{n-1}}$. Then the predictive kernel weight vector and covariance matrix are calculated. 

Specifically, the proposal particles at time $n-1$ are propagated through the transition function to calculate the \myComment{particles at time $n$,} represented as indigo points in Fig.~\ref{fig:04b}.  
%in Fig. 3(a2) as, 
\begin{align}
\label{eq:eqparticlesprediction}
 \mathbf{x}_n^{\{i\}} = f( \tilde{\mathbf{x}}_{n-1}^{\{i\}}, \mathbf{u}_n^{\{i\}}), \quad i = 1\dots M,  
\end{align}
where $\mathbf{u}_n^{\{i\}}$ represents a process noise sample drawn from the process noise \ac{pdf}.
The feature mappings of $ \mathbf{x}_n^{\{1:M\}}$ are  $\Phi_n = \left[\featurex(\mathbf{x}_n^{\{1\}}),\dots, \featurex(\mathbf{x}_n^{\{M\}})\right]$, 
and the predictive KME and covariance operator are calculated by
\begin{gather}
\label{eq:eqAKKF_prediction_mean}
\hat{\mu}^{-}_{\mathbf{x}_{n}}=\Phi_{n}\mathbf{w}^{-}_{n},\\
\label{eq:eqAKKF_prediction_Cov}
\hat{\mathcal{C}}^{-}_{\mathbf{x}_{n}\mathbf{x}_{n}}=\Phi_{n}\mathbf{S}^{-}_{n}\Phi_{n}^{\mathrm{T}}.
\end{gather}
Here, the weight vector $\mathbf{w}^{-}_{n}$ and matrix $\mathbf{S}^{-}_{n}$ are derived in \eqref{eq:eqAKKF_prediction1}–\eqref{eq:eqS_prediction} as follows. 
 The conditional KME of the transitional probability $\pdf(\mathbf{x}_n|\mathbf{x}_{n-1}, \mathbf{y}_{1:n-1})$ is approximated as
 \begin{equation}
 \label{eq:eqAKKF_prediction1}
\begin{aligned}
\pdf(\mathbf{x}_n|\mathbf{x}_{n-1}, \mathbf{y}_{1:n-1})   &\mapsto  \hat{\mu}_{\mathbf{x}_{n}}^{-} = \hat{\mathcal{C}}_{\mathbf{x}_{n}|\mathbf{x}_{n-1}}\hat{\mu}_{\mathbf{x}_{n-1}}^{+},%\\
    %&= \hat{\mathcal{C}}_{\mathbf{x}_{n} \mathbf{x}_{n-1}}\hat{\mathcal{C}}_{\mathbf{x}_{n-1} \mathbf{x}_{n-1}}^{-1}\hat{\mu}_{\mathbf{x}_{n-1}}^{+}.
\end{aligned}
 \end{equation}
where the empirical approximations to the
conditional embedding 
operator $\hat{\mathcal{C}}_{\mathbf{x}_{n}|\mathbf{x}_{n-1}}$ can be derived from a least-squares objective \cite{Grnewlder2012ConditionalME} as
%\begin{equation}
%\label{eq:eqAKKF_prediction2}
%  \begin{split}
 %    \hat{\mathcal{C}}_{\mathbf{x}_{n} \mathbf{x}_{n-1}} = \frac{1}{M}\sum_{i=1}^M \featurex(\mathbf{x}_{n}^{\{i\}})\otimes \featurex(\tilde{\mathbf{x}}_{n-1}^{\{i\}})
 %    =\frac{1}{M}\Phi_{n}\Psi^{\rm{T}}_{n-1},
%\end{split}
%    \end{equation}
%\begin{equation}
%\label{eq:eqAKKF_prediction3}
%  \begin{split}
%     \hat{\mathcal{C}}_{\mathbf{x}_{n-1} \mathbf{x}_{n-1}} = \frac{1}{M}\sum_{i=1}^M \featurex(\tilde{\mathbf{x}}_{n-1}^{\{i\}})\otimes \featurex(\tilde{\mathbf{x}}_{n-1}^{\{i\}})
 %    =\frac{1}{M}\Psi_{n-1}\Psi^{\rm{T}}_{n-1}.
%  \end{split}
%    \end{equation}
%Hence, the transition operator $\hat{\mathcal{C}}_{\mathbf{x}_{n}|\mathbf{x}_{n-1}}$ is given by
 \begin{equation}
 \label{eq:eqAKKF_prediction4}
  \begin{split}
\hat{\mathcal{C}}_{\mathbf{x}_{n}|\tilde{\mathbf{x}}_{n-1}}
    &%= \hat{\mathcal{C}}_{\mathbf{x}_{n} \tilde{\mathbf{x}}_{n-1}}\hat{\mathcal{C}}_{\tilde{\mathbf{x}}_{n-1} \tilde{\mathbf{x}}_{n-1}}^{-1} 
    = \Phi_{n}\left( \Psi_{n-1}\Psi^{\rm{T}}_{n-1} +\lambda_{\tilde{K}} I \right)^{-1}\Psi^{\rm{T}}_{n-1}\\
    &=\Phi_{n}\left( \GramxTT +\lambda_{\tilde{K}} I \right)^{-1}\Psi^{\rm{T}}_{n-1}.
     \end{split}
    \end{equation}
Substituting \eqref{eq:eqAKKF_proposal_weightmean}  and \eqref{eq:eqAKKF_prediction4} into \eqref{eq:eqAKKF_prediction1}, we have the  {estimate}  of the predictive empirical \ac{KME} of $\mathbf{x}_{n}$ as
  \begin{equation}
  \begin{split}
    &\hat{\mu}_{\mathbf{x}_{n}}^{-} = \hat{\mathcal{C}}_{\mathbf{x}_{n}|\tilde{\mathbf{x}}_{n-1}}\hat{\mu}_{\mathbf{x}_{n-1}}^{+}\\
    &=\Phi_{n} \left(\GramxTT
       + {\lambda_{\tilde{K}}}  I\right)^{-1}\Psi^{\rm{T}}_{n-1}
       \Phi_{n-1} \mathbf{w}_{n-1}^{+}\\
       & = \Phi_{n} \left(\GramxTT
       + {\lambda_{\tilde{K}}}  I\right)^{-1} \GramxT \mathbf{w}_{n-1}^{+}\\
       &= \Phi_{n}\mathbf{w}_{n}^{-}.
 \end{split}
    \end{equation}
 {Thus}, {the estimate of the predictive kernel weight vector is given by}
\begin{equation}
\label{eq:eqAKKF_prediction_weight}
\mathbf{w}^{-}_{n} = \left(\GramxTT
       + {\lambda_{\tilde{K}}}  I\right)^{-1} \GramxT\mathbf{w}_{n-1}^{+} = \Gamma_{n-1}  \mathbf{w}^{+}_{n-1}.
\end{equation}
From \eqref{eq:eqAKKF_proposal_1} and \eqref{eq:eqAKKF_prediction_weight},  we {see} that $\mathbf{w}^{-}_{n}  =\tilde{\mathbf{w}}^{+}_{n-1}$. 
Next, the empirical predictive covariance operator at time $n$ is computed as
\begin{equation}
\label{eq:eqAKKF_proposal_C}
 \begin{aligned}
 \hat{\mathcal{C}}_{\mathbf{x}_{n}\mathbf{x}_{n}}^{-}&= \hat{\mathcal{C}}_{\mathbf{x}_{n}|\tilde{\mathbf{x}}_{n-1}}
 \hat{\mathcal{C}}_{\mathbf{x}_{n-1}\mathbf{x}_{n-1}}^{+} \hat{\mathcal{C}}_{\mathbf{x}_{n}|\tilde{\mathbf{x}}_{n-1}}^{\mathrm{T}}+\mathcal{V}_{n}\\
 &= \hat{\mathcal{C}}_{\mathbf{x}_{n}|\tilde{\mathbf{x}}_{n-1}}
\Psi_{n-1} \tilde{S}_{n-1}^{+}\Psi_{n-1}^{\mathrm{T}}
\hat{\mathcal{C}}_{\mathbf{x}_{n}|\tilde{\mathbf{x}}_{n-1}}^{\mathrm{T}}+
\mathcal{V}_{n}\\
&= \Phi_n \tilde{S}_{n-1}^{+} \Phi_n^{\mathrm{T}} + \mathcal{V}_{n}.
\end{aligned} 
\end{equation}
Here,  $\mathcal{V}_{n}$ represents the  transition residual matrix, which is derived as
\begin{equation}
\label{eq:eqAKKF_proposal_V}
 \begin{aligned}
\mathcal{V}_{n} &= \frac{1}{M} \left( \hat{\mathcal{C}}_{\mathbf{x}_{n}|\tilde{\mathbf{x}}_{n-1}}\Psi_{n-1} - \Phi_{n} \right)\left( \hat{\mathcal{C}}_{\mathbf{x}_{n}|\tilde{\mathbf{x}}_{n-1}}\Psi_{n-1} - \Phi_{n} \right)^{\mathrm{T}}\\
= &\frac{1}{M} \left[ \Phi_{n} \left( \GramxTT+\lambda_{\tilde{K}} I \right)^{-1}
    \Psi_{n-1}^{\mathrm{T}}\Psi_{n-1} - \Phi_{n} \right]\\
    &\times \left[\Phi_{n}  \left( \GramxTT +\lambda_{\tilde{K}} I \right)^{-1}
    \Psi_{n-1}^{\mathrm{T}}\Psi_{n-1} - \Phi_{n} \right]^{\mathrm{T}}\\
    =&\Phi_{n} \frac{1}{M} \left[  \left( \GramxTT +\lambda_{\tilde{K}} I \right)^{-1}
   \GramxTT - I \right]\\
   &\times 
    \left[  \left( \GramxTT +\lambda_{\tilde{K}} I \right)^{-1}
   \GramxTT - I \right]^{\mathrm{T}}\Phi_{n}^{\mathrm{T}}\\
    \equiv &\Phi_{n} V_n \Phi_{n}^{\mathrm{T}}.
\end{aligned} 
\end{equation}
Here,  $ V_{n}$ is the finite matrix representation of $\mathcal{V}_{n} $. The predictive weight covariance matrix is given by substituting \eqref{eq:eqAKKF_proposal_C} and \eqref{eq:eqAKKF_proposal_V} into \eqref{eq:eqAKKF_prediction_Cov}; 
\begin{equation}
{S}_{n}^{-} = \tilde{S}_{n-1}^{+} +V_{n}.
\label{eq:eqS_prediction}
\end{equation}
%{where the transition residual  matrix  $V_{n}$ is calculated based on (25) while the Gram matrices are calculated based on the propagated particles but not the training data. and $\mathcal{V}_{n} = \Phi_n V_{n}\Phi_n^{\mathrm{T}}$.} 

\subsection{Update at Time $n$}

\myComment{This step modifies the predictive kernel weight vector and covariance matrix calculated in the previous step, considering the new observation at time $n$.}
The observation particles in Fig.~\ref{fig:041c} are updated based on the measurement model as
\begin{align}
  \mathbf{y}_n^{\{i\}} = h( {\mathbf{x}}_{n}^{\{i\}}, \mathbf{v}_n^{\{i\}}), \quad i = 1\dots M. 
\end{align}
%{where $v_n^{\{i\}}$ represent a measurement noise sample. } 
Here, $\mathbf{v}_n^{\{i\}}$ represents a measurement noise sample {drawn from the measurement noise \ac{pdf}}. 
Then,  the kernel mappings of observation particles in the kernel feature space  are $\Upsilon_n = \left[\featurey(\mathbf{y}_n^{\{1\}}),\dots, \featurey(\mathbf{y}_n^{\{M\}})\right]$. The posterior  KME is calculated as
\begin{gather}
\label{eq:eq:mu_update1}
    \hat{\mu}_{\mathbf{x}_n}^{+}=\hat{\mu}_{\mathbf{x}_n}^{-}
    +\mathcal{Q}_n
    \left[\featurey(\mathbf{y}_n)-\hat{\mathcal{C}}_{\mathbf{y}_n|\mathbf{x}_n}\hat{\mu}_{\mathbf{x}_n}^{-}\right],
\end{gather}
where  the kernel Kalman gain operator denoted as $\mathcal{Q}_n$ is applied to the correction term $\featurey(\mathbf{y}_n)-\hat{\mathcal{C}}_{\mathbf{y}_n|\mathbf{x}_n}\hat{\mu}_{\mathbf{x}_n}^{-}$ and is derived by minimizing the trace of the {posterior} covariance operator $\hat{\mathcal{C}}_{\mathbf{x}_n \mathbf{x}_n}^{+}$ \cite{Gebhardt}, as in the \eqref{eq:eqQ_n_main}:
\begin{equation}
\label{eq:eqQ_n_main}
    \begin{aligned}
    \mathcal{Q}_n &= \arg \min_{\mathcal{Q}_n} \mathrm{Tr}\left[{\hat{\mathcal{C}}_{\mathbf{x}_n \mathbf{x}_n}^{+}}\right]\\
    &=\Phi_n S_n^{-}  \left(\Gramy  S_n^{-}  + \kappa I  \right)^{-1}\Upsilon_n^{\mathrm{T}}.
    \end{aligned}
\end{equation}
The Appendix provides the derivation details of $\mathcal{Q}_n$. %\ref{appendix:a} 
Then, the updated \ac{KME} vector represented in \eqref{eq:eq:mu_update1} is calculated as
\begin{equation}
   \begin{aligned}
    \hat{\mu}_{\mathbf{x}_n}^{+}  = &\Phi_n {\mathbf{w}}^{+}_{n} 
     = \Phi_n {\mathbf{w}}^{-}_{n}
    +\mathcal{Q}_n
    \left[\phi(\mathbf{y}_n)-\hat{\mathcal{C}}_{\mathbf{y}_n|\mathbf{x}_n}\hat{\mu}_{\mathbf{x}_n}^{-}\right]\\
   = &  \Phi_n\left[  {\mathbf{w}}^{-}_{n}
   + S_n^{-}
   \left( \Gramy S_n^{-}  + \kappa I
\right)^{-1}\left(G_{:,\mathbf{y}_n}- \Gramy  {\mathbf{w}}^{-}_{n}
   \right)\right],
   \end{aligned} 
\end{equation}
where the kernel vector of the measurement at time $n$ is $G_{:,\mathbf{y}_n}=\Upsilon_n^{\mathrm{T}}\phi(\mathbf{y}_n)$, and
the Gram matrix of the observation at time $n$ is  $\Gramy  = \Upsilon^{\mathrm{T}}_n\Upsilon_n$. Hence, the weight vector is updated as
\begin{equation}
\label{eq:eqAKKF_update_w}
   \begin{aligned}
   {\mathbf{w}}^{+}_{n} &=   {\mathbf{w}}^{-}_{n}
   + S_n^{-}
   \left( \Gramy S_n^{-}  + \kappa I
\right)^{-1}\left(G_{:,\mathbf{y}_n}- \Gramy  {\mathbf{w}}^{-}_{n}
   \right)\\
   &={\mathbf{w}}^{-}_{n} +  Q_n\left(G_{:,\mathbf{y}_n}- \Gramy  {\mathbf{w}}^{-}_{n}
   \right),
\end{aligned} 
\end{equation}
where $Q_n$ is the finite matrix representation of $\mathcal{Q}_{n}$;
\begin{align}
\label{eq:eqAKKF_Q_N}
Q_n = S_n^{-}
   \left(\Gramy  S_n^{-} + \kappa I
\right)^{-1}.
\end{align}
Then, the covariance operator can be expressed as{:}
\begin{equation}
\label{eq:eqAKKF_update_Cxx2}
   \begin{aligned}
   \hat{\mathcal{C}}_{\mathbf{x}_n \mathbf{x}_n}^{+} = \hat{\mathcal{C}}_{\mathbf{x}_n \mathbf{x}_n}^{-} - \mathcal{Q}_n \Upsilon_n 
   S_n^{-}\Phi_n^{\mathrm{T}}.
\end{aligned} 
\end{equation}
The {derivation details are shown in Appendix.} As the predictive and posterior covariance operators are
   $\hat{\mathcal{C}}_{\mathbf{x}_n \mathbf{x}_n}^{-} = \Phi_n S_n^{-}\Phi_n^{\mathrm{T}}$ and 
     $\hat{\mathcal{C}}_{\mathbf{x}_n \mathbf{x}_n}^{+} = \Phi_n S_n^{+}\Phi_n^{\mathrm{T}}$, \eqref{eq:eqAKKF_update_Cxx2} is rewritten as
\begin{equation}
   \begin{aligned}
   &\Phi_n S_n^{+}\Phi_n^{\mathrm{T}} = \Phi_n S_n^{-}\Phi_n^{\mathrm{T}} - \mathcal{Q}_n \Upsilon_n 
   S_n^{-}\Phi_n^{\mathrm{T}}\\
   \Rightarrow 
&\Phi_n S_n^{+}\Phi_n^{\mathrm{T}}\\
&= \Phi_n S_n^{-}\Phi_n^{\mathrm{T}} - \Phi_n S_n^{-}\Upsilon_n^{\mathrm{T}}
\left( \Upsilon_n  S_n^{-} \Upsilon_n^{\mathrm{T}} + \kappa I
\right)^{-1} \Upsilon_n 
   S_n^{-}\Phi_n^{\mathrm{T}}\\
  &=  \Phi_n \left [S_n^{-} - S_n^{-}
  \Upsilon_n^{\mathrm{T}}
\left( \Upsilon_n  S_n^{-} \Upsilon_n^{\mathrm{T}} + \kappa I
\right)^{-1} \Upsilon_n S_n^{-}\right]\Phi_n^{\mathrm{T}}.
\end{aligned} 
\end{equation}
Therefore, the kernel weight covariance matrix is finally updated as
\begin{equation}
\label{eq:eqAKKF_update_S}
   \begin{aligned}
  S_n^{+} &= S_n^{-} - S_n^{-}
  \Upsilon_n^{\mathrm{T}}
\left( \Upsilon_n  S_n^{-} \Upsilon_n^{\mathrm{T}} + \kappa I
\right)^{-1} \Upsilon_n S_n^{-}\\
&= S_n^{-} - 
S_n^{-}
   \left( \Gramy  S_n^{-}  + \kappa I
\right)^{-1} \Gramy 
S_n^{-}\\
&= S_n^{-} - Q_n \Gramy 
S_n^{-}.
\end{aligned} 
\end{equation}
%\ikpc{I need to check to see if this involves assuming $\Upsilon_n^{-1}$ exists.}

\subsection{Implementation of AKKF }

\myComment{Based on the above descriptions, Algorithm \ref{alg:algorithm1} summarizes the implementation of the AKKF.}
%\ikpc{It seems to me that $K_{xx}$ in step 3 is not needed. At least it is nor clear where it would be used.}
\begin{algorithm}
    \caption{Adaptive kernel Kalman filter }
  \begin{algorithmic}[1]
  \label{alg:algorithm1}
    \REQUIRE DSSM: transition model $f(\cdot)$ and measurement model $h(\cdot)$.
    \STATE \textbf{Initialization}: Set the initial particles in real space $\mathbf{x}_0^{\{i=1:M\}} \sim P_{\text{init}}$,
    $\Phi_0: = \left[\featurex(\mathbf{x}_0^{\{1\}}), \dots,\featurex(\mathbf{x}_0^{\{M\}}) \right]$, $\mathbf{w}_0 = 1/M \left[1,\dots,1\right]^{\rm{T}}$,
     $\Psi_0 = \Phi_0$,  $\Gamma_0 = I$.
    \FOR{$n = 1 : N$}
    
      \STATE Prediction:
      %\vspace{-4mm}
      %\begin{multicols}{2}
\begin{itemize}
          \item First, in the data space:  
          
          $\mathbf{x}_n^{\{i\}} = f( \tilde{\mathbf{x}}_{n-1}^{\{i\}},\mathbf{u}_n^{\{i\}}) $, $i = 1\dots M$.
          %\begin{align*}
          % x_n^{\{i\}} = f( \tilde{x}_{n-1}^{\{i\}}), \quad i = 1\dots M.   
          %\end{align*}
      \item[{$\Rightarrow$}] Second, in the kernel feature space:
      
      $\Phi_n = \left[\featurex(\mathbf{x}_n^{\{1\}}),\dots, \featurex(\mathbf{x}_n^{\{M\}})\right]$, %$\Gramx = \Phi_n^{\rm{T}}\Phi_n$.
      
{$\mathbf{w}^{-}_{n} =  \tilde{\mathbf{w}}^{+}_{n-1}$,}

${S}_{n}^{-} = \tilde{S}_{n-1}^{+} +V_{n}$.
      %\begin{align*}
       %   \mathbf{w}^{-}_{n} &= T_{n-1} \mathbf{w}^{+}_{n-1},\\
       %   {S}_{n}^{-} &= \tilde{S}_{n-1}^{+} +V_n.
      %\end{align*}
    %  \begin{equation}
   %\begin{aligned}
  % V_n 
  %  =& \frac{1}{M} \left[\left(\Psi_{n-1}^{\rm{T}}\Psi_{n-1}+\lambda I\right)^{-1}\Psi_{n-1}^{\rm{T}}\Phi_n -I\right],\\
  % &\times 
  % \left[\left(\Psi_{n-1}^{\rm{T}}\Psi_{n-1}+\lambda I\right)^{-1}\Psi_{n-1}^{\rm{T}}\Phi_n -I\right]^{\rm{T}}.
%\end{aligned} 
%\end{equation}
\end{itemize}
%\end{multicols}
%\vspace{-4mm}
  \STATE Update:
 % \vspace{-4mm}
      %\begin{multicols}{2}
         \begin{itemize}
    \item First, in the data space:
    
    $ \mathbf{y}_n^{\{i\}} = h( \mathbf{x}_{n}^{\{i\}}, \mathbf{v}_{n}^{\{i\}})$, $i = 1\dots M$.
          %\begin{gather*}
          %     y_n^{\{i\}} = h( \tilde{x}_{n}^{\{i\}}), \quad i = 1\dots M. 
          %\end{gather*}
         \item[{$\Rightarrow$}]  Second, in the kernel feature space: 
         
          $\Upsilon_n = \left[\featurey(\mathbf{y}_n^{\{1\}}),\dots, \featurey(\mathbf{y}_n^{\{M\}})\right]$, $\Gramy = \Upsilon_n ^{\rm{T}}\Upsilon_n $.
        
         ${\mathbf{w}}^{+}_{n} = {\mathbf{w}}^{-}_{n} +  Q_n\left(G_{:,\mathbf{y}_n}- \Gramy  {\mathbf{w}}^{-}_{n}
   \right)$.
   
     $S_n^{+} =S_n^{-} - Q_n \Gramy  S_n^{-}.$

The  posterior KME with  the statistical information:

$\hat{\mu}_{\mathbf{x}_n} = \Phi_n \mathbf{w}^{+}_{n} =  \left[\mathbb{E}{\left(\mathbf{x}_n\mathbf{x}_n^{\rm{T}}\right)},
\mathbb{E}{\left(\mathbf{x}_n\right)}, c\right]^{\rm{T}}$.
        
 \end{itemize}
  %    \end{multicols}
      %\vspace{-4mm}
     \STATE Proposal particles draw:
       %\vspace{-4mm}
        % \begin{multicols}{2}
        \begin{itemize}
           %\item[{$\Leftarrow$}]  
          \item First, 
           in the data space:

          $\tilde{\mathbf{x}}_{n}^{\{i=1:M\}} \sim \mathcal{N}\left( \mathbb{E}\left(\mathbf{x}_n\right),  \mathbb{E}\left(\mathbf{x}_n\mathbf{x}_n^{\rm{T}}\right)-\mathbb{E}\left(\mathbf{x}_n\right)\mathbb{E}\left(\mathbf{x}_n\right)^{\rm{T}}  \right)$.

    \item[{$\Rightarrow$}]  Second, in kernel feature space:
    
    $\Psi_n = \left[\featurex(\tilde{\mathbf{x}}_{n}^{\{1\}}), \dots, \featurex(\tilde{\mathbf{x}}_{n}^{\{M\}}) \right]$. 
    
    $\Gamma_n = \left(\Psi_n^{\rm{T}}\Psi_n+\lambda I \right)^{-1} {\Psi_n^{\rm{T}}\Phi_n}$. 
     
     $\tilde{\mathbf{w}}^{+}_{n} = \Gamma_{n} \mathbf{w}^{+}_{n}$.
    
     $\tilde{\mathbf{S}}^{+}_{n} = \Gamma_{n} \mathbf{S}^{+}_{n} \Gamma_{n}^{\rm{T}}$.
  
       \end{itemize}
       
       %\end{multicols}  
    \ENDFOR
  \end{algorithmic}
\end{algorithm}

\section{Simulation Results}

\myComment{We report on three numerical examples showing the benefits of the proposed AKKF when the system \acp{DSSM} are available. In the first experiment, we deal with the state estimation problem following the \ac{UNGM}. We employ the \ac{UNGM} because of its high non-linearity and bimodality. 
We then report the tracking performance for the nonlinear-in-observation 
\acf{BOT} model, which is of interest in defense applications, with the target moving following either the linear \acf{CV} model or non-linear \ac{CT} model with an unknown and random walk turn rate, respectively. We compare the most commonly used state-of-the-art model-based filters,  i.e., the \ac{UKF}, \ac{GPF}, and bootstrap \ac{PF}.
%This paper mainly investigates proposing an efficient Bayesian filter to realize real-time tracking or estimation. The existing kernel-based filters, e.g., \acf{KKR}, require training data and an offline training process, which is invalid without training data or dramatically declined with improper training data. Therefore, we compare the proposed AKKF to the commonly used model-based Bayesian filters, i.e., the \ac{UKF}, \ac{GPF}, and \ac{PF}. 
}

 %\subsection{Synthetic Data Experiments}
%To generate synthetic data, we used the following DSSM, 
%\begin{align}
%    x_n &= \cos (2 \pi x_{n-1}) + u_n,\\
%    y_n &= \ex\pdf(x_n) + v_n.
%\end{align}
%$x_0 = 0.5$, and $\pdf(x_0)  \sim \mathcal{N}(0.5,1)$. The process noise and measurement noise are $u_n \sim \mathcal{N}(0,1) $ and $v_n \sim \mathcal{N}(0,1)$.

\subsection{State Estimation under UNGM}
%\ikp{The \ac{UNGM} is chosen due to the high non-linearity and bimodal in nature.}{}
%\ikpc{This is just a repetition of the previous paragraph.} 
The 
\ac{DSSM}  of \ac{UNGM} is written as \cite{978374}
\begin{align}
    x_n &= \alpha x_{n-1} + \beta \frac{x_{n-1}}{ 1+  x_{n-1}^2} + \gamma \cos\left(1.2 \left(n-1\right)\right) + u_n,\\
    y_n &= \frac{x_{n}^2}{ 20} + v_n.
\end{align}
Here, the process noise $ u_n$ and measurement noise $v_n$ are \acp{AWGN}, \myComment{i.e.,} $ u_n \sim \mathcal{N}(0,\sigma_u^2)$, and $ v_n \sim \mathcal{N}(0,\sigma_v^2)$.  We set
$x_0 = 0.1$, $\sigma_u^2 = 1$, $\sigma_v^2 = 1$. 
$\alpha = 0.5$, $\beta = 25$, $\gamma = 8$  \cite{978374}. 
The data sequence length is set to be $N = 100$.
We compare the estimation performance of the \myComment{proposed \ac{AKKF} using a quadratic kernel} with the \ac{GPF}, and the bootstrap \ac{PF}, 
based on the following \ac{MSE} metric:
\begin{align}
    \mathrm{MSE} = \frac{1}{N} \sum_{n=1}^{N}(x_n-\hat{x}_n)^2.
\end{align}

\begin{figure}[!t]
	\centering
	\includegraphics [width=8cm,height=4.6cm]{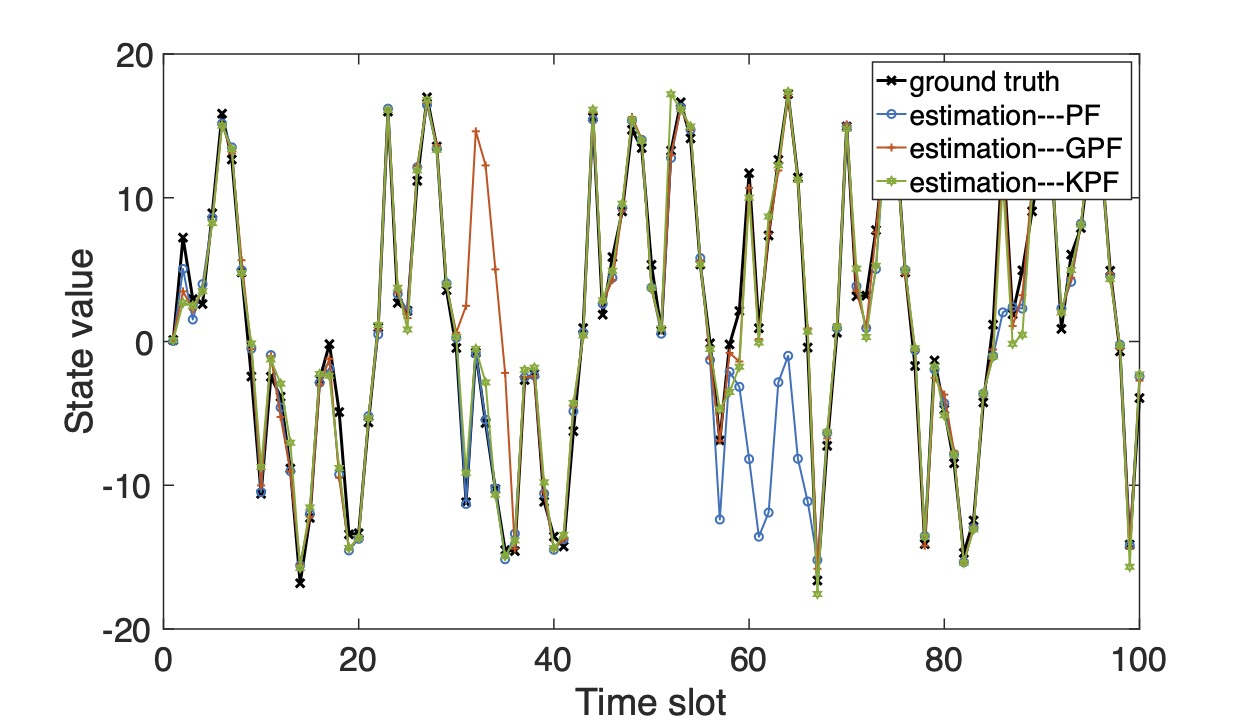}
	\caption{{ {Estimation performance comparison  of the PF,  GPF, and AKKF filters for the \ac{UNGM} in 100 time slot, the number of particles is set as $M=20$.} }}
	\label{fig:figUNGM1}
\end{figure}

\begin{figure}[!t]
	\centering
	\includegraphics [width=7.2cm,height=5cm]{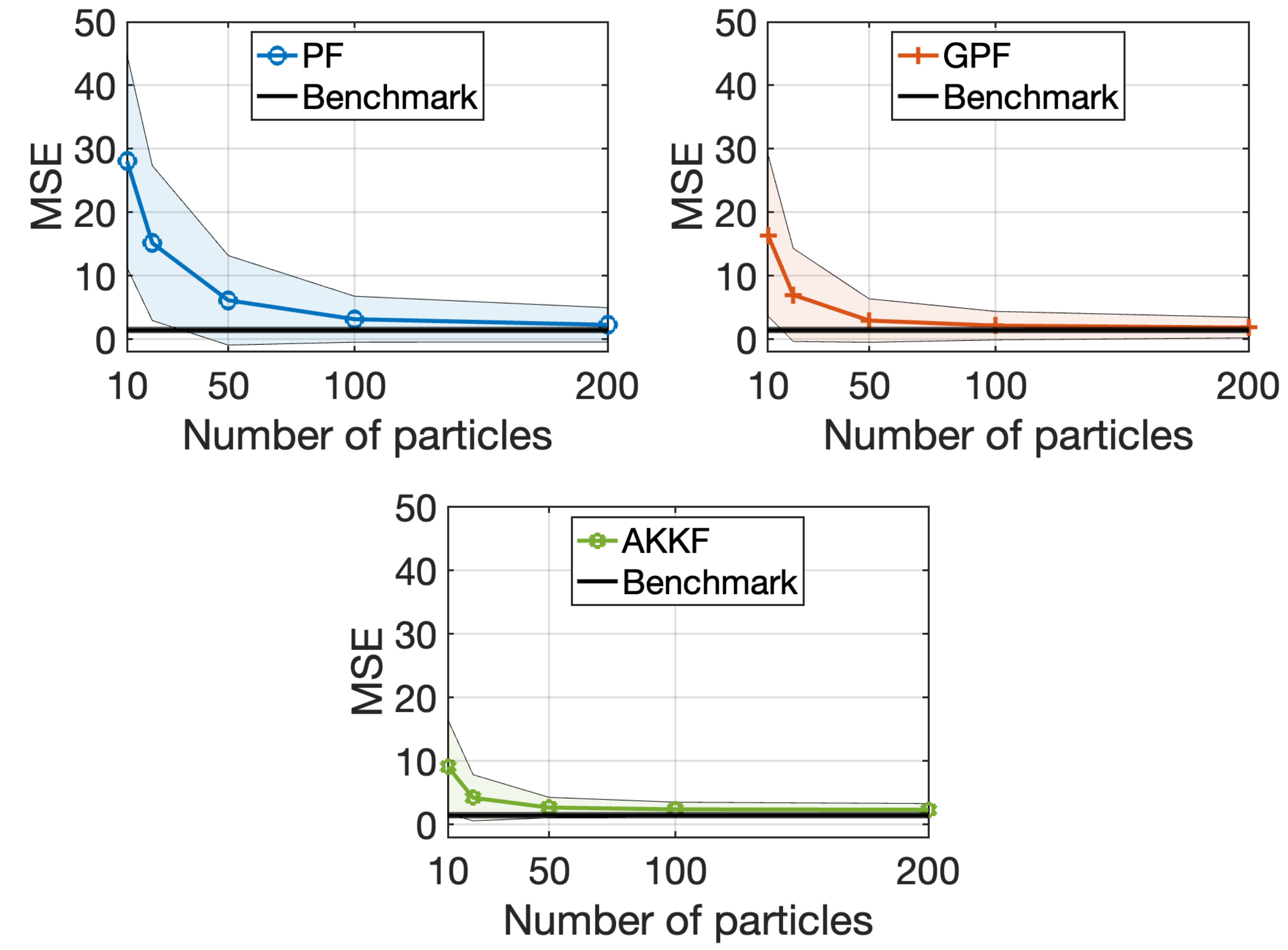}
	\caption {Performance comparison of the PF,  GPF, and AKKF filters with  an increasing  number of particles.
	%The number of \ac{MC} random realizations is set to be 1000. 
	Legend: Solid lines are the average MSEover 1000 random  \ac{MC} realizations, 
 \myComment{i.e.,} $\mathbb{E}(\mathrm{MSE})$; The colored areas are error bars calculated as $\mathbb{E}(\mathrm{MSE}) \pm \mathrm{Std}(\mathrm{MSE})$.} 
	\label{fig:figUNGM12}
\end{figure}

We compare the three filters through two simulations.  First, in Fig.~\ref{fig:figUNGM1}, we show the
states and the estimates obtained using filters with {$M = 20$ particles} for a single realization. From Fig.~\ref{fig:figUNGM1}, \myComment{the proposed \ac{AKKF} shows improved estimation performance compared with the bootstrap PF and the GPF which fail to track the ground truth state at specific points.}
%\ikpc{We should say why figure 5 leads us to this conclusion \myComment{i.e.,} both the GPF and the PF loose track but the AKKF does not.}
%\ikpc{Above we say "As expected" but I don't think it was necessarily so. Rather figure 5 tells us that the AKKF is better. Hence I'd delete "As expected".}
Fig.~\ref{fig:figUNGM12} shows the \ac{MSE} for 1000  random \acf{MC} realizations with the increasing {number of particles} $M=\left[10,20,50,100,200\right]$.
The benchmark performance is achieved by the bootstrap \ac{PF} with 2000 particles. From 
Fig.~\ref{fig:figUNGM12}, we can conclude that for the state estimation under the \ac{UNGM}, the proposed \ac{AKKF} shows \myComment{a distinct advantage} for a small number of particles, \myComment{i.e.,} $M = \left[10,20,50\right]$.


%\begin{figure}
 %   \centering
 %   \subfigure[] {\label{fig:06a}
%			\includegraphics[width=6cm,height = 4cm]{UNGM20-PF.jpg}}
%			\subfigure[] {\label{fig:06b}
%			\includegraphics[width=6cm,height = 4cm]{UNGM20-GPF.jpg}}
%			\subfigure[] {\label{fig:06c}
%			\includegraphics[width=6cm,height = 4cm]{UNGM20-AKKF.jpg}}
%    \caption{Performance comparison of the PF,  GPF, and AKKF filters with  increasing  particle  numbers.
%	The number of \ac{MC} random realizations is set to be 1000. 
%	Legend: Solid lines are the average over 1000 random  \ac{MC} realizations, , \myComment{i.e.,} $\mathbb{E}(\mathrm{MSE})$ the colored areas are error bars calculated as. $\mathbb{E}(\mathrm{MSE}) \pm \mathrm{Std}(\mathrm{MSE})$}
%	\label{fig:figUNGM12}
%\end{figure}

\subsection{\myComment{Bearing-only Tracking (BOT) – Linear Motion Behavior}}\label{sssec:num1}

The \ac{BOT} problem is of interest for airborne radar and sonar in passive listening mode and electronic warfare systems \cite{1232326}. \myComment{This paper considers the BOT problem with one object moving in a 2-D space.} The hidden state $\mathbf{x}_n = [\xi_n, \dot{\xi}_n,  \eta_n, \dot{	\eta}_n]^{T}$, where  $(\xi_n,\eta_n)$ and $(\dot{\xi}_n,\dot{\eta}_n)$ represent the target position and the corresponding velocity on X-axis and Y-axis, respectively. The moving trajectory is assumed to follow a \ac{CV} motion model, which is represented as
\begin{equation}
    \mathbf{x}_n = \mathbf{F} \mathbf{x}_{n-1}  + \mathbf{G} \mathbf{u}_n, \; \; n = 1,\dots,N,
\end{equation}
where $N=30$, the process noise is a $2\times 1$ vector, \myComment{i.e.,} $\mathbf{u}_n =\left[ u_x,u_y \right]_n^{\mathrm{T}}$, which follows a Gaussian distribution $\mathbf{u}_n \sim
\mathcal{N} (\mathbf{0}, \sigma^2_u \mathbf{I}_2)$, $\sigma_u =  1e^{-3}$ and $\mathbf{I}_2$ is the $2\times 2$ identity matrix. 
\begin{align*}
    \mathbf{F} = \begin{bmatrix}
1 & T_s & 0 & 0\\
0 & 1 & 0 & 0\\
0 & 0 & 1 & T_s \\
0 & 0 & 0 & 1\\
\end{bmatrix}, \;\;\;
  \mathbf{G} = \begin{bmatrix}
0.5 & 0  \\
1 & 0  \\
0 & 0.5  \\
0 & 1  \\
\end{bmatrix}
\end{align*}
\myComment{where $T_s$ is the sampling interval and is set as $T_s=1$.} The  prior distribution for the initial state  is specified as $\mathbf{x}_0 \sim \mathcal{N}(\bar{\mathbf{x}}_0, \mathbf{P}_0)$. Following \cite{1232326}, {we set the parameters of the prior distribution to be}
  $\bar{\mathbf{x}}_0 = \left[ -0.05, 0.001, 0.7, -0.05\right]^{\mathrm{T}}$ and
\begin{align*}
    \mathbf{P}_0 = \begin{bmatrix}
0.1 & 0 & 0 & 0\\
0 & 0.005 & 0 & 0\\
0 & 0 & 0.1 & 1\\
0 & 0 & 0 & 0.01\\
\end{bmatrix}.
\end{align*}

Although the motion model in this example is linear, the measurement model is non-linear, leading to non-Gaussian state distributions. We model the measurements as the actual  bearing with an additional Gaussian error term,
\begin{align}
    y_n = \tan^{-1}(\frac{\eta_n}{\xi_n}) + v_n.
    \label{eq:eqBOTmeasurement}
\end{align}
Here, the inverse tangent is the four-quadrant inverse tangent function,  %$v_n$ is an \ac{AWGN},  
$v_n\sim \mathcal{N}(0,\sigma^2_v)$,  $\sigma_v = 5e^{-3}$. 

\subsubsection{Tracking performance}

\begin{figure}
    \centering
    \subfigure[] {\label{fig:One_realisation_tracking1}
			\includegraphics[width=7.5cm,height = 5cm]{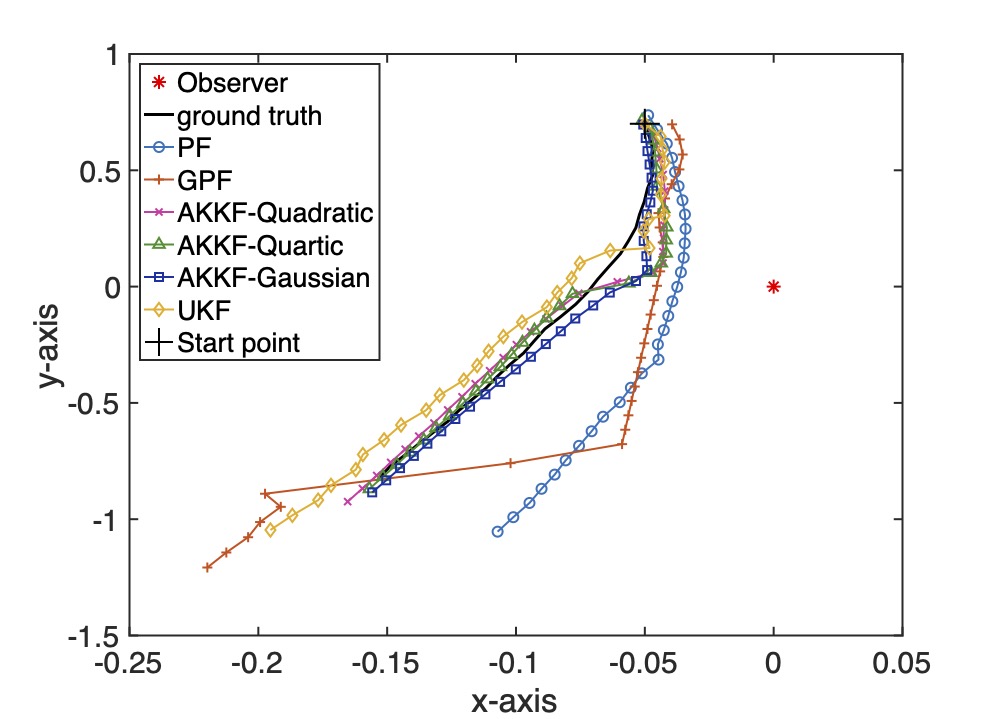}}
			\subfigure[] {\label{fig:One_realisation_tracking2}
			\includegraphics[width=7.5cm,height = 5cm]{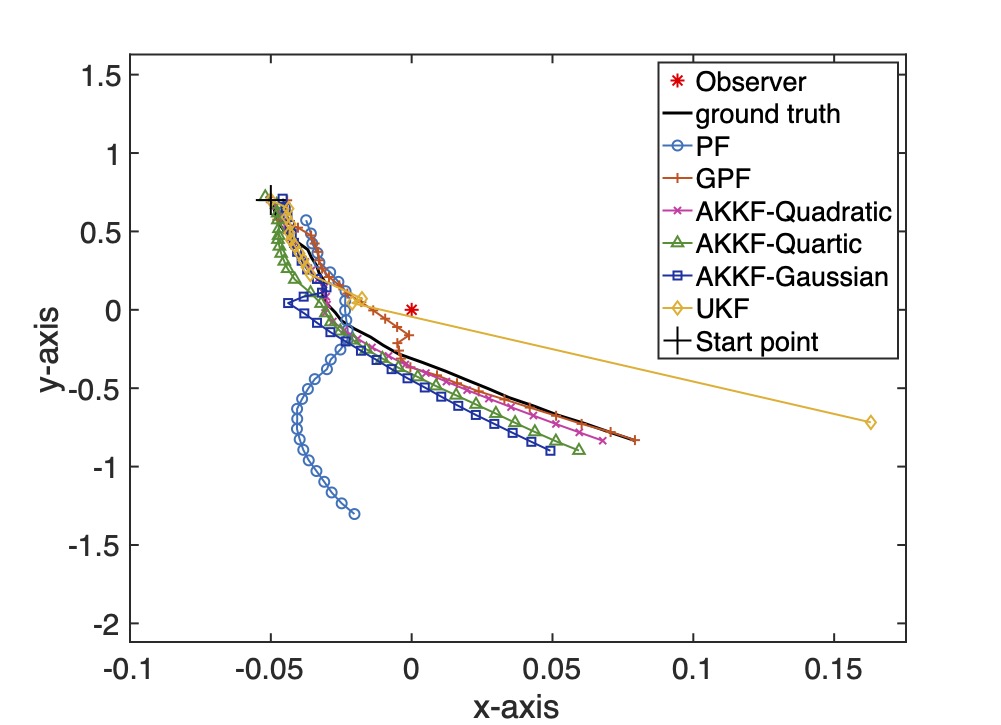}}
    \caption{BOT performance  of a moving target in two dimensions with UKF,  GPF, PF, quadratic kernel-based AKKF, quartic kernel-based AKKF, and Gaussian kernel-based AKKF. {The number of particles} for GPF, PF, and \acp{AKKF} is $M = 20$. Legend: $*$: the observer, $+$: the start point of moving trajectory. (a) Trajectory-1, (b) Trajectory-2.}
    \label{fig:tracking_realization}
\end{figure}

Fig.~\ref{fig:tracking_realization} 
 displays two representative trajectories and the tracking performance obtained by six filters: {UKF, GPF, PF, the proposed AKKF using finite quadratic kernel and quartic kernels,} and the proposed AKKF using infinite Gaussian kernel. We locate the observer at $\left[0, 0\right]$. The number of particles used for the \ac{PF}, GPF, and \acp{AKKF} is 20. The number of sigma points  for the UKF is 19.
 It can be seen from Fig.~\ref{fig:tracking_realization} that with a small {number of particles}, {divergence} may occur for the \ac{PF}, GPF, and UKF, while divergence is not observed for the proposed \acp{AKKF}.
%\begin{figure*}
%   \label{tab:my_label}
%    \centering
%    \includegraphics[width=16cm,height=9cm]{Image/lMSE_Comparison_random1000.pdf}
%    \caption{LMSE performance obtained by the \ac{UKF}, \ac{PF}, GPF, quadratic kernel based AKKF, quartic  kernel based AKKF, and Gaussian kernel based AKKF with increasing particle numbers. Legend: Solid lines are the average value of \acp{LMSE}, \myComment{i.e.,} $\mathbb{E}(\mathrm{LMSE})$ for 1000 random \ac{MC} realizations, colored areas are error bars $\mathbb{E}(\mathrm{LMSE}) \pm \mathrm{Std}(\mathrm{LMSE})$. }
 %   \label{fig:lMSE_Comparison}
%\end{figure*}
\begin{figure}[!t]
	\centering
	\includegraphics [width=8cm,height=6.5cm]{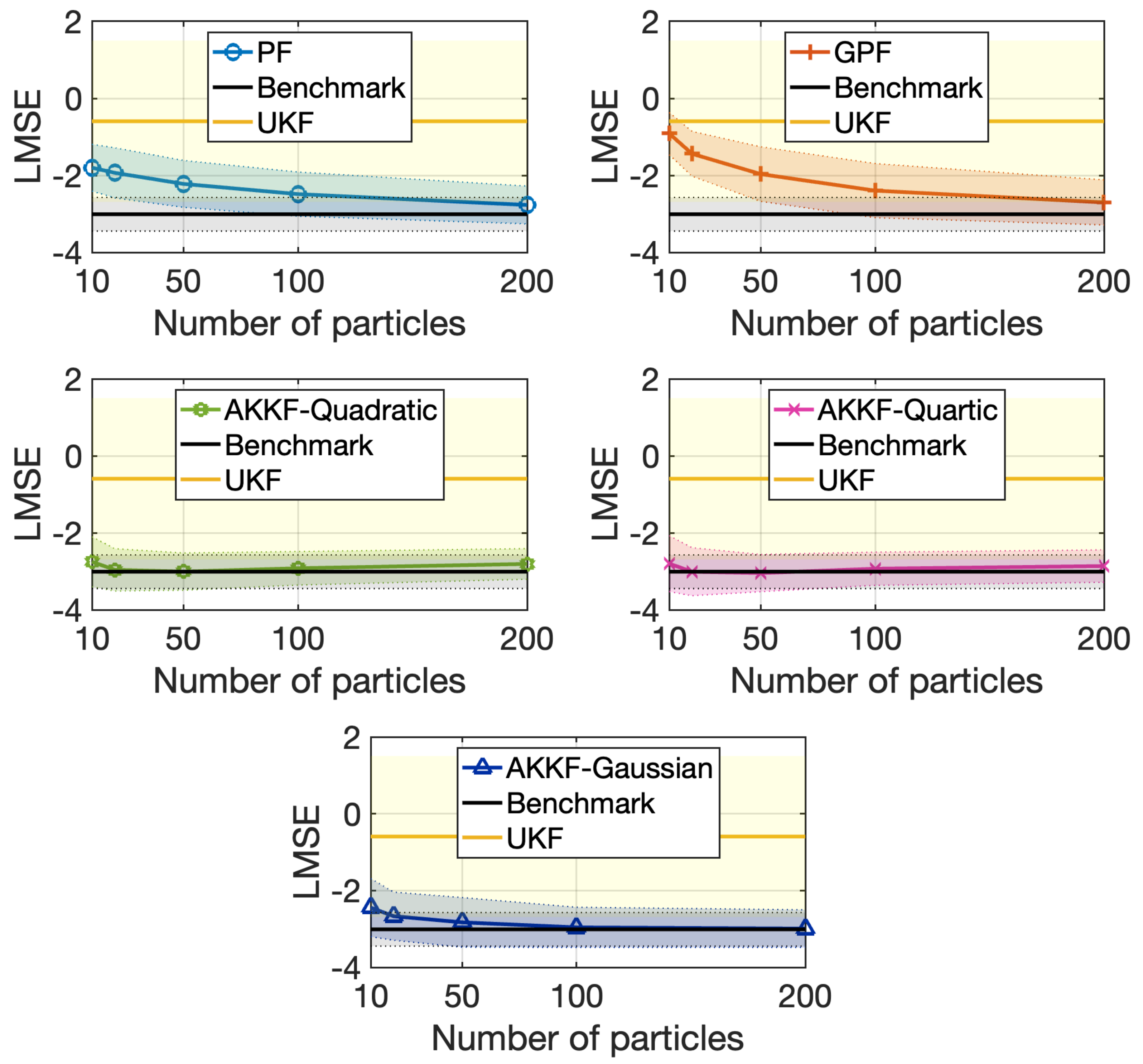}
	\caption{LMSE performance obtained by the \ac{UKF}, \ac{PF}, GPF, quadratic kernel-based \ac{AKKF}, quartic  kernel-based \ac{AKKF}, and Gaussian kernel-based \ac{AKKF} with an increasing number of particles. Legend: Solid lines are the average value of \acp{LMSE}, \myComment{i.e.,} $\mathbb{E}(\mathrm{LMSE})$ for 1000 random \ac{MC} realizations. The colored areas are error bars $\mathbb{E}(\mathrm{LMSE}) \pm \mathrm{Std}(\mathrm{LMSE})$.}
	\label{fig:lMSE_Comparison}
\end{figure}

%\begin{figure}
%   \label{tab:my_label}
%    \centering
%    \subfigure[] {\label{fig:04c}
%			\includegraphics[width=5.7cm,height = 4cm]{LMSE_PF.jpg}}
%			\subfigure[] {\label{fig:04c}
%			\includegraphics[width=5.7cm,height = 4cm]{LMSE_GPF.jpg}}
%			\subfigure[] {\label{fig:04c}
%			\includegraphics[width=5.7cm,height = 4cm]{LMSE_AKKF_2.jpg}}
%			\subfigure[] {\label{fig:04c}
%			\includegraphics[width=5.7cm,height = 4cm]{LMSE_AKKF_4.jpg}}
%			\subfigure[] {\label{fig:04c}
%			\includegraphics[width=5.7cm,height = 4cm]{LMSE_AKKF_G.jpg}}
%    \caption{LMSE performance obtained by the \ac{UKF}, \ac{PF}, GPF, quadratic kernel-based \ac{AKKF}, quartic  kernel-based \ac{AKKF}, and Gaussian kernel-based \ac{AKKF} with an increasing number of particles. Legend: Solid lines are the average value of \acp{LMSE}, \myComment{i.e.,} $\mathbb{E}(\mathrm{LMSE})$ for 1000 random \ac{MC} realizations. The colored areas are error bars $\mathbb{E}(\mathrm{LMSE}) \pm \mathrm{Std}(\mathrm{LMSE})$.}
 %   \label{fig:lMSE_Comparison}
%\end{figure}

Fig.~\ref{fig:lMSE_Comparison} shows the average  \acf{LMSE} obtained for 1000 random \ac{MC} realizations for all the position state variables. The \ac{LMSE} is defined as,
 \begin{align}
\label{eq:eqLMSE}
    \mathrm{LMSE} = \log\left[\frac{1}{N} \sum_{n=1}^{N}\sqrt{(\xi_n-\hat{\xi}_n)^2 +(\eta_n-\hat{\eta}_n)^2 }\right].
\end{align}
The numbers of particles are set to be $M = \left[10, 20, 50, 100, 200\right] $. The compared filters {are} the PF, the GPF, and the \acp{AKKF} using quadratic kernel, quartic kernel, and Gaussian kernel, respectively. The benchmark performance is achieved by the bootstrap \ac{PF} with {$10^4$} particles. From Fig.~\ref{fig:lMSE_Comparison}, we arrive at the following conclusions. 
First, the proposed \acp{AKKF} show significant improvement compared to the \ac{PF} and \ac{GPF} with the same number of particles, especially with small numbers of particles, \myComment{i.e.,} $M = \left[10,20,50\right]$. Second, on average, the \ac{AKKF} using the quartic kernel performs better than the \ac{AKKF} using the quadratic kernel. \myComment{The improved performance is likely due to the quartic feature mappings incorporating more statistical information about the hidden state. The \acp{AKKF} using quadratic and quartic kernels can approach the benchmark performance with 20 particles.
It is interesting that the \ac{LMSE} performance slightly deteriorates here as the number of particles increases. This appears to be caused by the overuse of particles, which is likely to lead to singular or badly scaled Gram matrices, increasing the inaccuracy of matrix inversion. Hence, estimation biases propagate to reduce the tracking performance. }

\begin{figure}[!t]
	\centering
	\includegraphics [width=9cm,height=5.6cm]{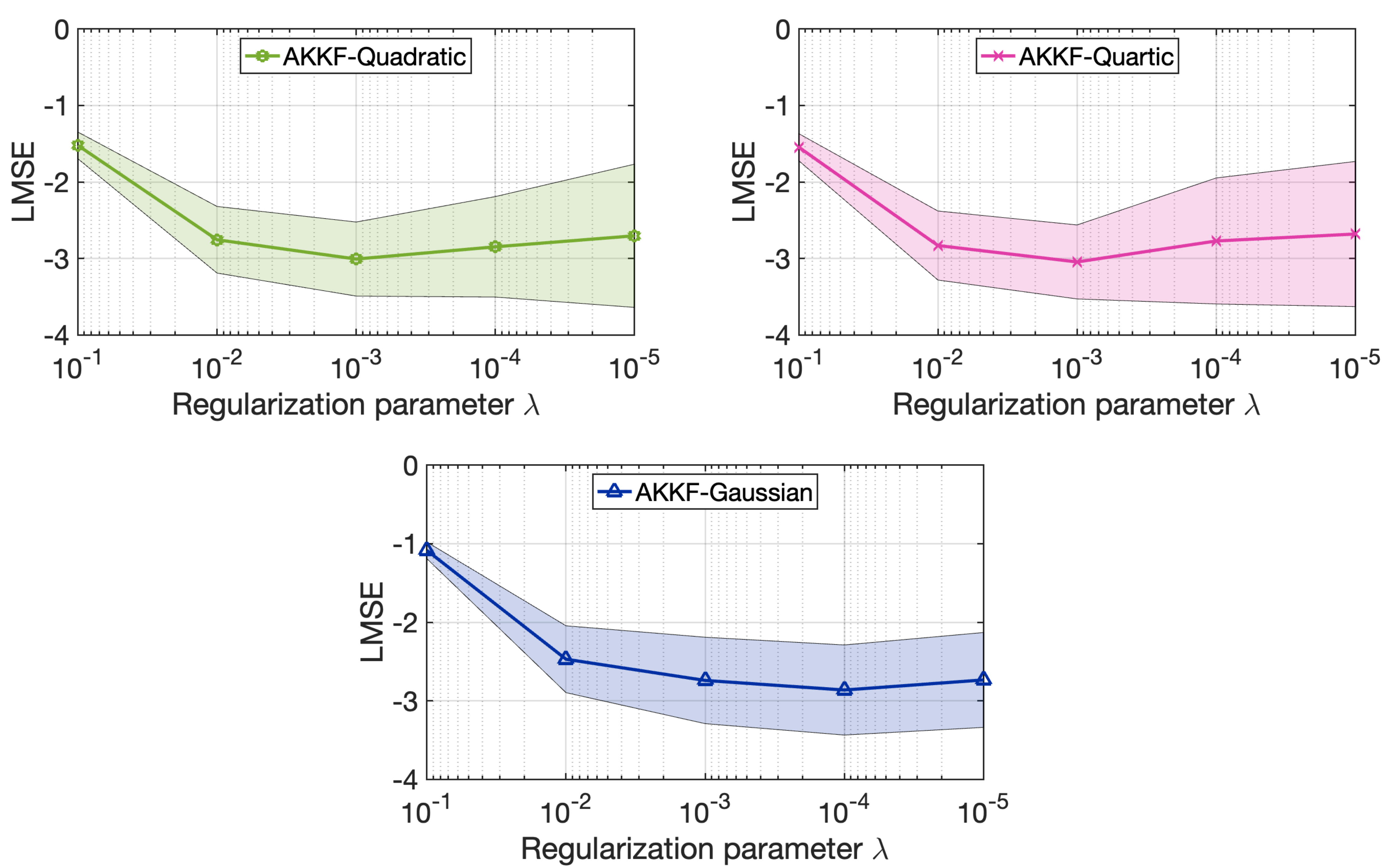}
	\caption{LMSE performance comparison of the PF,  GPF, and AKKF filters with  the varying regularization parameter $\lambda$.
	The number of \ac{MC} random realizations is set to 1000.
	Legend: Solid lines are the average over 1000 random \ac{MC} realizations,  \myComment{i.e.,} $\mathbb{E}(\mathrm{MSE})$. The colored areas are error bars calculated as  $\mathbb{E}(\mathrm{MSE}) \pm \mathrm{Std}(\mathrm{MSE})$.}
	\label{fig:lMSE_Comparison_lambda}
\end{figure}

Next, we investigate the effects of varying regularization  parameters $\lambda$ and $\kappa$ 
on the tracking performance. The former is used in the calculation of the transition matrix $\Gamma_n$ in \eqref{eq:eqAKKF_proposal_2}. The latter is used for the calculation of kernel Kalman gain $Q_n$ in \eqref{eq:eqAKKF_Q_N}. \myComment{The regularization parameter choice must be derived from the real data. Hence, we investigate the good empirical value of regularization parameters using an \ac{MC} method.} In this simulation, $\kappa$ is set to be equal to $\lambda$, and the number of particles for AKKF is set to be 50.
From Fig.~\ref{fig:lMSE_Comparison_lambda}, we can see that {the LMSE performance is relatively insensitive to the values of $\lambda$ and $\kappa$ when they are in the range} $\left[10^{-4}, 10^{-2}\right]$.

%\begin{figure}
%   \label{fig:figLambda}
%    \centering
%    \subfigure[] {\label{fig:04c}
%			\includegraphics[width=7.1cm,height = 4.7cm]{LMSE_effect_lambda_AKKF_2.jpg}}
%			\subfigure[] {\label{fig:04c}
%			\includegraphics[width=7.1cm,height = 4.7cm]{LMSE_effect_lambda_AKKF_4.jpg}}
%			\subfigure[] {\label{fig:04c}
%			\includegraphics[width=7.1cm,height = 4.7cm]{LMSE_effect_lambda_AKKF_G.jpg}}
%    \caption{LMSE performance comparison of the PF,  GPF, and AKKF filters with  the varying regularization parameter $\lambda$.
%	The number of \ac{MC} random realizations is set to 1000. 
%	Legend: Solid lines are the average over 1000 random \ac{MC} realizations,  \myComment{i.e.,} $\mathbb{E}(\mathrm{MSE})$. The colored areas are error bars calculated as  $\mathbb{E}(\mathrm{MSE}) \pm \mathrm{Std}(\mathrm{MSE})$.}
%    \label{fig:lMSE_Comparison_lambda}
%\end{figure}

\begin{figure}
    \centering
    \subfigure[] {\label{fig:04c}
			\includegraphics[width=7.5cm,height = 3.8cm]{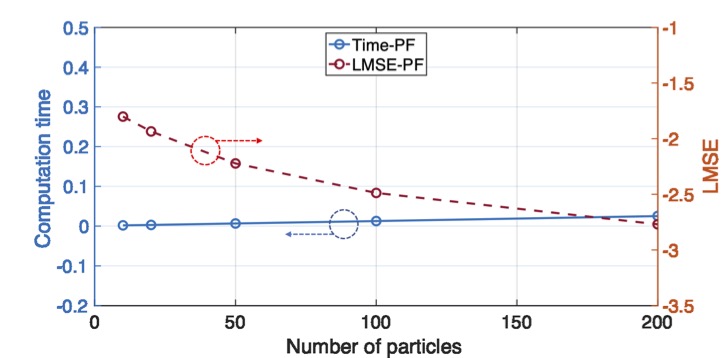}}
			\subfigure[] {\label{fig:04c}
			\includegraphics[width=7.5cm,height = 3.8cm]{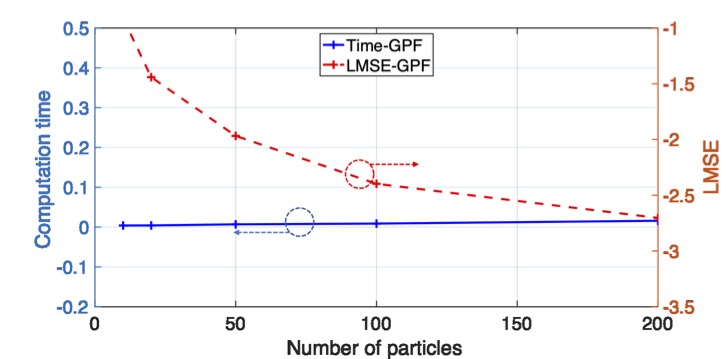}}
			\subfigure[] {\label{fig:04c}
			\includegraphics[width=7.5cm,height = 3.8cm]{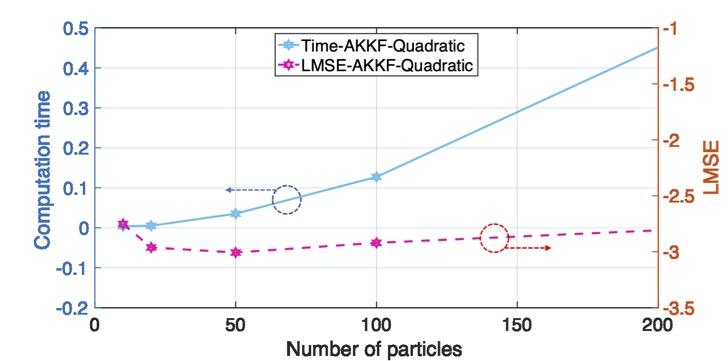}}
			\subfigure[] {\label{fig:04c}
			\includegraphics[width=7.5cm,height = 3.8cm]{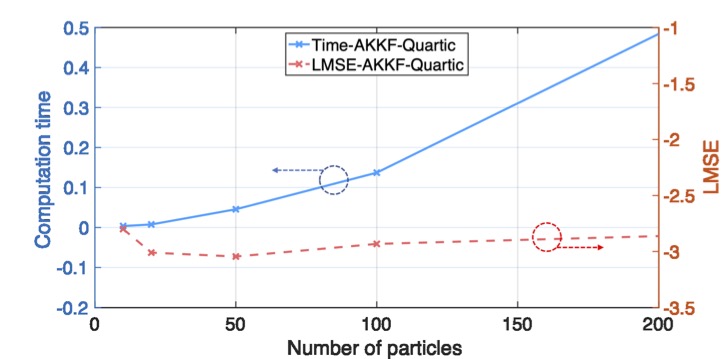}}
			\subfigure[] {\label{fig:04c}
			\includegraphics[width=7.5cm,height = 3.8cm]{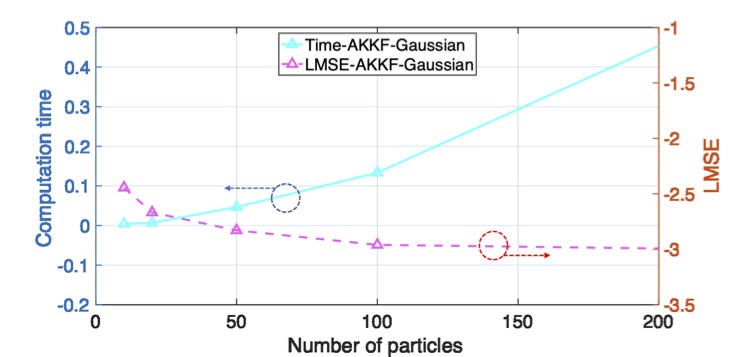}}
    \caption{\myComment{Showing the trend of the computation time and \ac{LMSE} of the  \ac{PF}, \ac{GPF}, and different AKKFs; Average computation time (s) and \ac{LMSE} over 1000 \ac{MC} random realizations with increasing number of particles. Legend: Blue circles with arrows mean that the curves are the performance of computation time; Red circles with arrows mean that the curves show the LMSE performance. }}%The GPF and PF follow a linear trend when the number of particles $10 \leq M \leq 200$, while the \acp{AKKF} follows a quadratic trend. 
    \label{fig:figComputatiom_time_Number_particles}
\end{figure}
\begin{figure}[!t]
\centering
    \subfigure[] {\label{fig:04c_1}
			\includegraphics[width=6.2cm,height = 2.8cm]{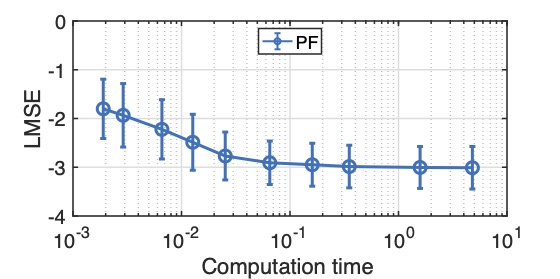}}
			\subfigure[] {\label{fig:042_2}
			\includegraphics[width=6.2cm,height = 2.8cm]{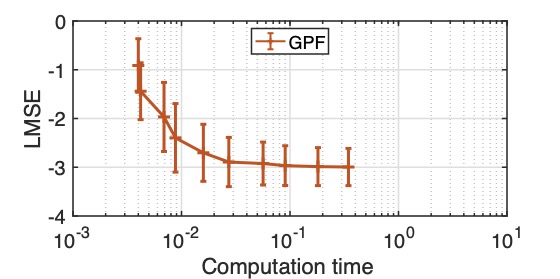}}
			\subfigure[] {\label{fig:04c_3}
			\includegraphics[width=6.2cm,height = 2.8cm]{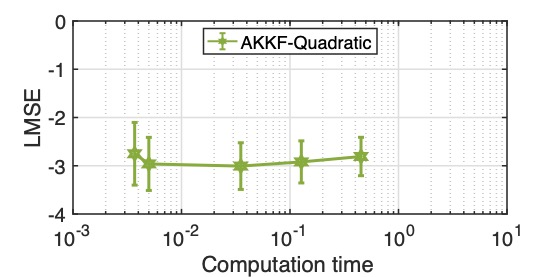}}
			\subfigure[] {\label{fig:04c_4}
			\includegraphics[width=6.2cm,height = 2.8cm]{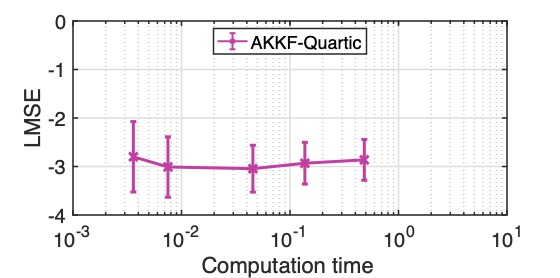}}
			\subfigure[] {\label{fig:04c_5}
			\includegraphics[width=6.2cm,height = 2.8cm]{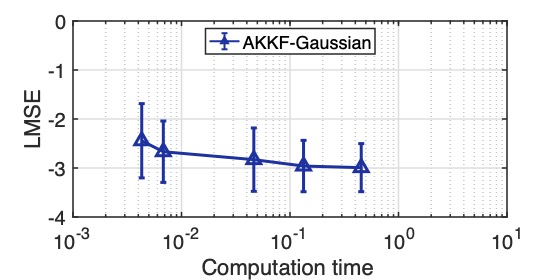}}
	\caption{\myComment{Computation time and \ac{LMSE}; Average computation time (s) and  \ac{LMSE} over 1000 \ac{MC} random realizations.}}
	\label{fig:figComputatiom_time_LMSE.pdf}
\end{figure}

\subsubsection{Computational complexity} 
\myComment{In the next experiment, we compare the computation time of filters and show the results 
in Fig.~\ref{fig:figComputatiom_time_Number_particles}.} The simulations are implemented in Matlab and run using MacBook Pro, Chip Apple M1. Fig.~\ref{fig:figComputatiom_time_Number_particles} shows the average computation time obtained for 1000 \ac{MC} realizations, from which we can {see} that the computation time of the bootstrap \ac{PF} increases linearly with the increase of particle numbers, while the computation time  of the proposed \ac{AKKF} increases quadratically  with the increase of particle numbers $M$ when $10 \leq M \leq  200$, since the computational complexity of matrices inversion increases quadratically. Even though the increasing trend of computational complexity for  the \ac{AKKF} is more significant, the \ac{LMSE} tracking performance of the \ac{AKKF} can approach the benchmark, \myComment{e.g.,} $-3.0$, with very small number of particles requirement. For a further {confirmation} of this conclusion, Fig.~\ref{fig:figComputatiom_time_LMSE.pdf} shows the \ac{LMSE} performance with the correspond running time. {From this figure, we can  conclude that with the \ac{LMSE} performance benchmark is $-3.0$, the computation time for the \ac{PF}, \ac{GPF}, quadratic kernel-based \ac{AKKF}, quartic kernel-based \ac{AKKF} and Gaussian kernel-based \ac{AKKF} are 0.35s, 0.35s, 0.035s, 0.0075s and 0.45s, respectively. }

\begin{figure}[!t]
	\centering
	\includegraphics [width=8cm,height=7.5cm]{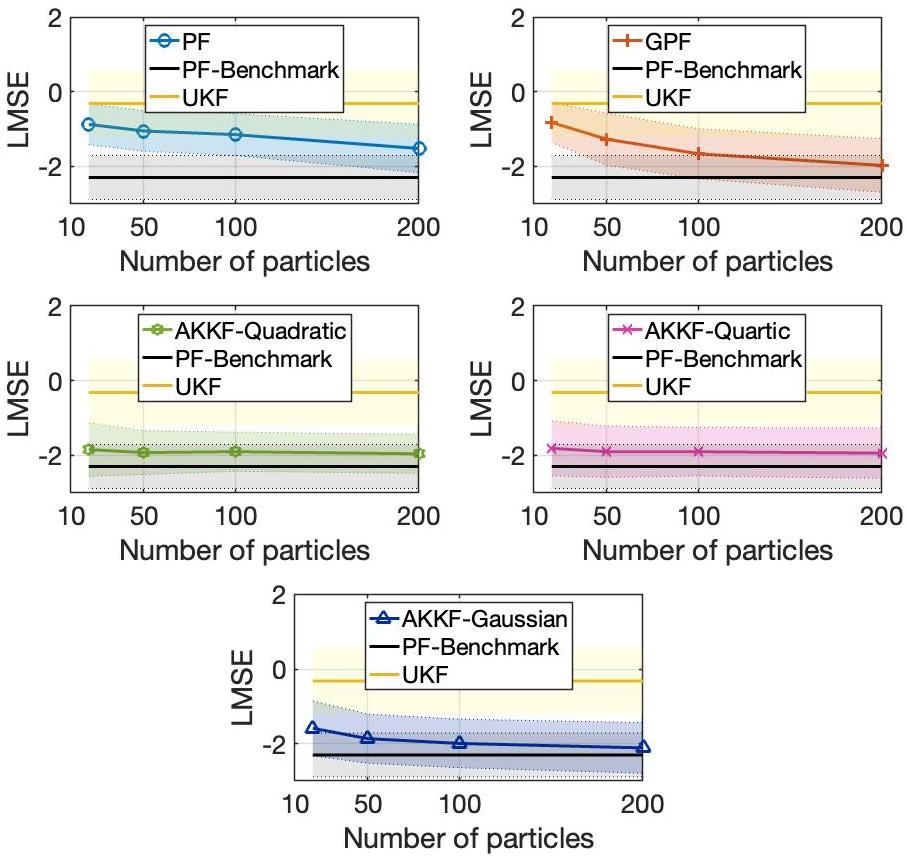}
	\caption{\myComment{LMSE performance for the BOT tracking under the \ac{CT} model with unknown and random walk turn rate. Legend: Solid lines are the average value of \acp{LMSE}, \myComment{i.e.,} $\mathbb{E}(\mathrm{LMSE})$ for 1000 random \ac{MC} realizations. The colored areas are error bars $\mathbb{E}(\mathrm{LMSE}) \pm \mathrm{Std}(\mathrm{LMSE})$.}}
	\label{fig:lMSE_Comparison_CT}
\end{figure}

\myComment{\subsection{Bearing-only Tracking (BOT) – Highly Maneuvering Behaviors}
In our final experiment, we consider the same BOT observation model with a nonlinear motion model. 
 The motion behavior of hidden states $\mathbf{x}_{n} = \left[\xi_{n}, \dot{\xi}_{n},  \eta_{n}, \dot{\eta}_{n}, \omega_n \right]^{\mathrm{T}}, n = 1,\dots,N$ is set to follow \ac{CT} model with unknown and dynamic turn rate as \cite{1261132}
 \begin{equation}
 \begin{split}
    %\mathbf{H_2}: \quad 
     \begin{bmatrix}  \xi_{n}\\\dot{\xi}_{n}\\ \eta_{n}\\
     \dot{\eta}_{n} \end{bmatrix} =
    \begin{bmatrix}
  & 1 & \frac{\sin \omega_{n-1}T_s }{\omega_{n-1}} &0 & -\frac{1-\cos \omega_{n-1}T_s }{\omega_{n-1}} \\ 
  & 0 & \cos \omega_{n-1}T_s & 0 &-\sin \omega_{n-1}T_s \\
  & 0 &\frac{1-\cos \omega_{n-1}T_s }{\omega_{n-1}} & 1 & \frac{\sin \omega_{n-1}T_s }{\omega_{n-1}} \\
  & 0 & \sin \omega_{n-1}T_s & 0 & \cos \omega_{n-1}T_s\\
    \end{bmatrix} \mathbf{x}_{n-1}   + \mathbf{v}_{n},
    \label{eq:eqH2}
    \end{split}
\end{equation}
\begin{equation}
\omega_n = \begin{cases}
    \omega_{n-1} + v_{n,\omega},& \text{if }  n \neq N/2\\
    \omega_{n-1}/3 + v_{n,\omega},              & \text{otherwise}
\end{cases}
\end{equation}
where $\omega_n$ is the random walk turn rate and changes at $n=N/2$. The sampling interval is set as $T_s = 1$, $\mathbf{v}_{n} \sim \mathcal{N}(\mathbf{0}, \sigma^2_v R)$, $v_{n,\omega} \sim \mathcal{N}({0}, \sigma^2_{w})$, $\sigma_v = 1e^{-3}$, and $\sigma_\omega = 1e^{-2}$,
%. The process noise $\mathbf{v}_{n}$ and turn rate noise $v_{n,\omega}$ follow \acp{AWGN}, as   $\mathbf{v}_{n} \sim \mathcal{N}(\mathbf{0}, \sigma^2_v R)$ and $v_{n,\omega} \sim \mathcal{N}({0}, \sigma^2_{w})$, respectively. %where  $\sigma^2_v$ represents power spectral density of the \ac{AWGN}.
%In the simulation  we set $\sigma_v = 1e^{-3}$, $\sigma_\omega = 1e^{-2}$; 
 \begin{equation*}
\begin{split}
    & R=\\ 
  & \begin{bmatrix}
  & \frac{2(\omega_{n}T_s - \sin \omega_{n}T_s)}{\omega_{n}^3} & \frac{1- \cos \omega_{n}T_s}{\omega_{n}^2}  & 0 & \frac{\omega_{n}T_s - \sin \omega_{n}T_s}{\omega_{n}^2}   \\ 
  & \frac{1 - \cos \omega_{n}T_s}{\omega_{n}^2}& T_s & -\frac{\omega_{n}T_s - \sin \omega_{n}T_s}{\omega_{n}^3} & 0    \\
  & 0 &-\frac{\omega_{n}T_s - \sin \omega_{n}T_s}{\omega_{n}^3} & \frac{2(\omega_{n}T_s - \sin \omega_{n}T_s)}{\omega_{n}^3}  &\frac{1- \cos \omega_{n}T_s}{\omega_{n}^2} \\
  & \frac{\omega_{n}T_s - \sin \omega_{n}T_s}{\omega_{n}^2}  & 0 & \frac{1- \cos \omega_{n}T_s}{\omega_{n}^2} & T_s 
    \end{bmatrix}. 
\end{split}
\end{equation*}
The initial position and velocity states' prior distribution follows the settings in Section \ref{sssec:num1}.  The  prior distribution for the unknown turn rate is $ \omega_0 \sim \mathcal{U} \left[ 0, \pi/6 \right]$. 
 The measurement model and corresponding settings follow \eqref{eq:eqBOTmeasurement}.  
 \begin{figure}
    \centering
    \subfigure[] {\label{fig:One_realisation_tracking1_CT}
			\includegraphics[width=8cm,height = 5.5cm]{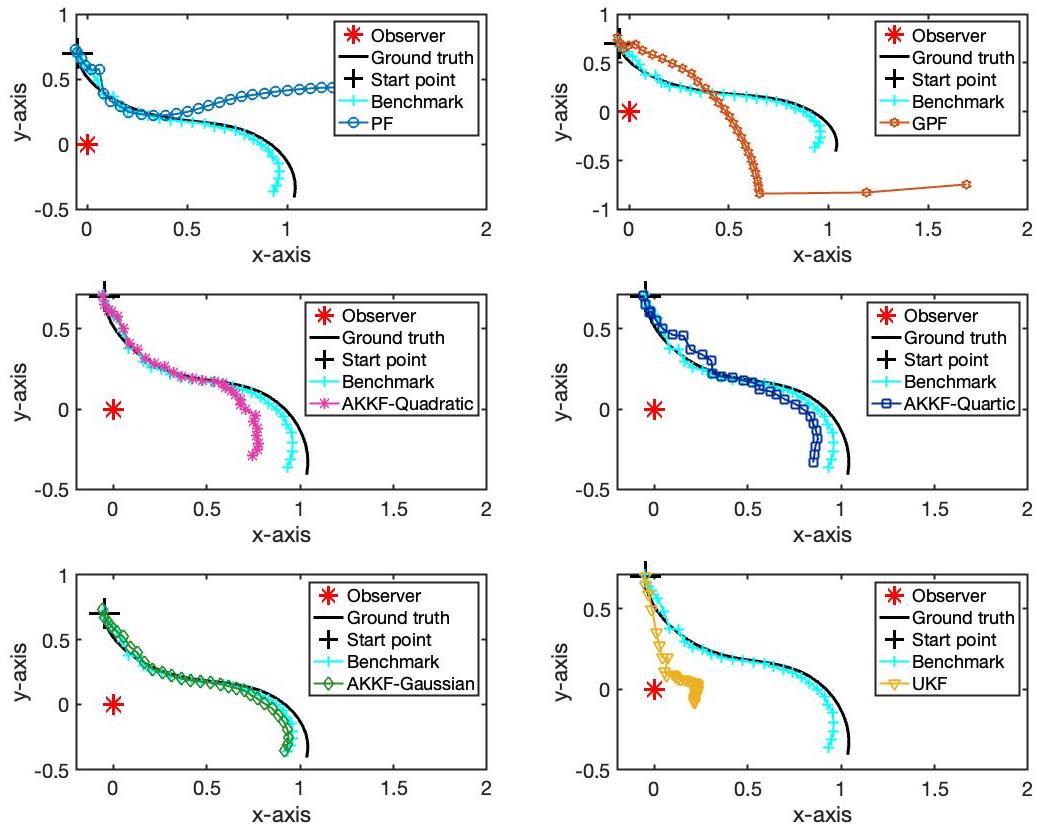}}
			\subfigure[] {\label{fig:One_realisation_tracking2_CT}
			\includegraphics[width=8cm,height = 5.5cm]{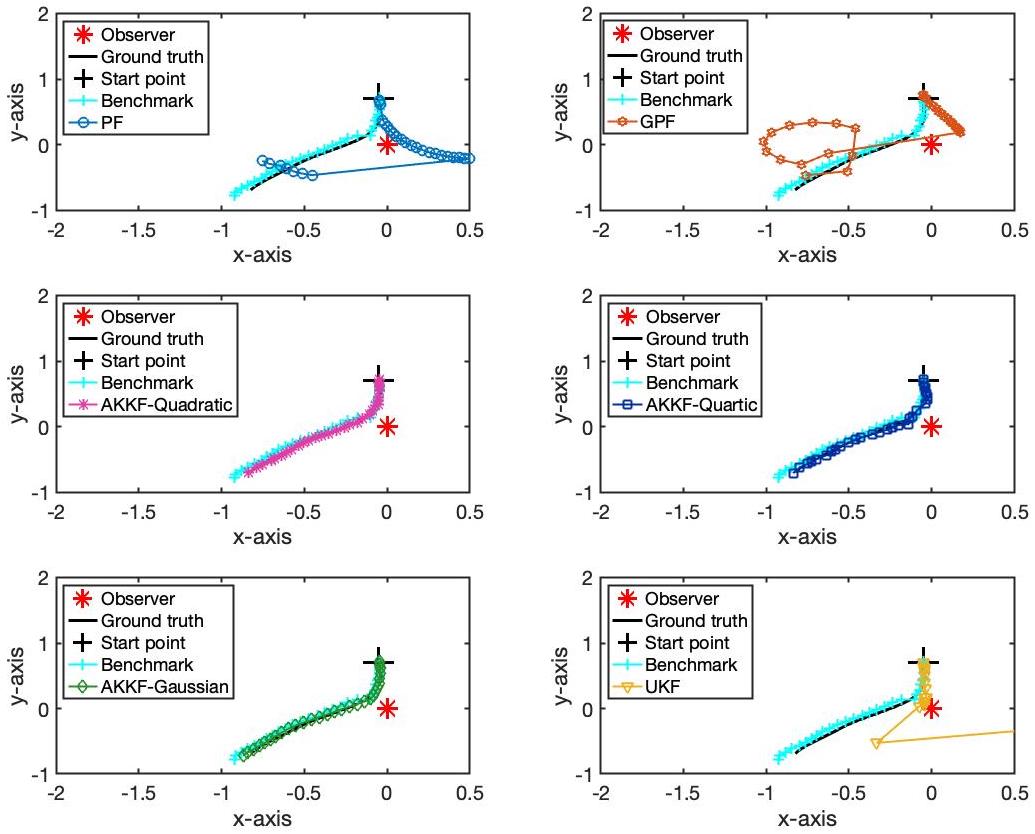}}
    \caption{\myComment{Tracking performance of a moving target following \ac{CT} model with random walk turn rate. % with UKF,  GPF, PF, quadratic kernel-based AKKF, quartic kernel-based AKKF, and Gaussian kernel-based AKKF.
    The number of particles for GPF, PF, and \acp{AKKF} is $M = 100$. Legend: $*$: the observer, $+$: the start point of moving trajectory. (a) Trajectory-1, (b) Trajectory-2.}}
    \label{fig:tracking_realization_CA}
\end{figure}

Fig.~\ref{fig:tracking_realization_CA} 
 displays two representative trajectories and the tracking performance obtained by six filters: UKF, GPF, PF, the AKKF with quadratic, quartic and Gaussian kernels. The PF with $10^4$ particles is used as a benchmark. The number of particles used for the compared \ac{PF}, GPF, and AKKFs is 100. {The number of sigma points  for the UKF is 23.}
 Fig.~\ref{fig:lMSE_Comparison_CT} shows the average  \ac{LMSE} obtained for 1000 random \ac{MC} realizations for all the position state variables. We set the numbers of particles as $M = \left[20, 50, 100, 200\right] $.  From Fig.~\ref{fig:tracking_realization_CA} and Fig.~\ref{fig:lMSE_Comparison_CT},
 we  conclude that divergence is  more severe for the \ac{PF}, \ac{GPF}, and \ac{UKF} than the proposed \acp{AKKF}, and
the proposed \acp{AKKF} still significantly improve performance with small numbers of particles when the target is undergoing non-linear motion behavior. However, the performance of the \acp{AKKF} with quadratic and quartic kernels can't be enhanced with the increased number of particles when  $M>50$. This appears to be caused by the fact that quadratic and quartic kernels only allow modeling features of data up to the order of the polynomial, but for the \ac{BOT} systems in which the target behaves following highly maneuvering, quadratic and quartic kernels are not effective enough to capture the diversity of the non-linearities.}

\section{Conclusions}

In this paper, we provided a new approach to model-driven
Bayesian filters. By embedding the predictive and posterior
\acp{pdf} into \acp{RKHS}, classical \ac{KF} calculation can be employed along with an adaptive sampling of the \ac{DSSM} to predict the new data space information. We have observed that more feature information
of the hidden states and the observations can be captured and
recorded with a significantly smaller number of particles than are needed in \ac{PF}-based methods while retaining equivalent estimation accuracy. Furthermore, as the new filters are comprised of standard matrix-vector multiplication operations, the overall
computational complexity is also very favorable and offers an excellent opportunity for parallelization.

%In this paper, we provided a new approach to model-driven Bayesian filters. By embedding the predictive and posterior \acp{pdf} into kernel statistical spaces,  more feature information of the hidden states and {the }observations can be captured and recorded \jh{with smaller number of particles.} 
%Therefore, the {number of particles employed can be} reduced significantly {whilst still retaining equivalent} estimation accuracy. %{\color{red}{Furthermore, the choice of the RKHS allows the user to select the appropriate complexity for the given task.}In the tasks considered here, this proposed filter serves to regularize the filter solution and {\color{red}therefore enables superior performance with fewer particles than in the \ac{PF} and \ac{GPF}.}
%Furthermore as the new filters are comprised of standard matrix vector multiplication operations, the overall computationalcomplexity is also very favorable.

%(if you do a compute time comparison you can possibly say more here)

%\newpage
\appendix
%\begin{appendices}

%\subsection{Kernel Kalman gain and updated kernel covariance operator}
%\label{appendix:a}

This Appendix gives the derivations  of kernel Kalman gain $\mathcal{Q}_n$ and updated kernel covariance operator $\hat{\mathcal{C}}_{x_n x_n}^{+}$ that follow \cite{Gebhardt} but are included here for completeness.

The trace of the posteriori covariance operator $\hat{\mathcal{C}}_{\mathbf{x}_n \mathbf{x}_n}^{+}$  is defined as  $\hat{\mathcal{C}}_{\mathbf{x}_n \mathbf{x}_n}^{+} = \mathbb{E}\left[\epsilon_n \epsilon_n^{\rm{T}}\right]$, where $\epsilon_n$ is the error of the posteriori \ac{KME} and calculated as
\begin{equation}
  \begin{aligned}
  \epsilon_n &= \featurex(\mathbf{x}_n)-\hat{\mu}_{\mathbf{x}_n}^{+}\\
    &= \featurex(\mathbf{x}_n)-\left[\hat{\mu}_{\mathbf{x}_n}^{-}
    +\mathcal{Q}_n
    \left(\featurey(\mathbf{y}_n)- \hat{\mathcal{C}}_{\mathbf{y}_n|\mathbf{x}_n}\hat{\mu}_{\mathbf{x}_n}^{-}-\mathcal{R}\right)
   \right]\\
   &= \left(I -  \mathcal{Q}_n\hat{\mathcal{C}}_{\mathbf{y}_n|\mathbf{x}_n}\right)
   \left( \featurex(
  \mathbf{x}_n)-\hat{\mu}_{\mathbf{x}_n}^{-}\right)
   -\mathcal{Q}_n \mathcal{R},
  \end{aligned}
  \end{equation}
  where we have used the fact that $ \featurey(\mathbf{y}_n) = \mathcal{C}_{\mathbf{y}_n|\mathbf{x}_n} \featurex(
  \mathbf{x}_n)$.
 Then, noting that $\featurex(\mathbf{x}_{n-1})-\hat{\mu}_{\mathbf{x}_{n-1}}^{+} = \epsilon_{n-1}$, $\hat{\mathcal{C}}_{\mathbf{x}_n \mathbf{x}_n}^{+} $ is calculated as
\begin{equation}
\label{eq:eqAKKF_update_Cxx}
   \begin{aligned}
   \hat{\mathcal{C}}_{\mathbf{x}_n\mathbf{x}_n}^{+}%&= \mathbb{E}\left[\epsilon \epsilon^{\rm{T}}\right]
   =\left(I -  \mathcal{Q}_n\hat{\mathcal{C}}_{\mathbf{y}_n|\mathbf{x}_n}\right)
   \hat{\mathcal{C}}_{\mathbf{x}_n \mathbf{x}_n}^{-}
   \left(I -  \mathcal{Q}_n\hat{\mathcal{C}}_{\mathbf{y}_n|\mathbf{x}_n}\right)^{\mathrm{T}} + \mathcal{Q}_n \mathcal{R} \mathcal{Q}_n^{\mathrm{T}},
\end{aligned} 
\end{equation}
where $\mathcal{R}$ is the covariance matrix of the residual of the observation operator.
The trace of $\hat{\mathcal{C}}_{\mathbf{x}_n\mathbf{x}_n}^{+}$ is minimized when its matrix derivative with respect to the gain matrix is zero.
\begin{equation}
\label{eq:eqAKKF_update_Derivete}
\begin{aligned}
    &\frac{\partial \hat{\mathcal{C}}_{\mathbf{x}_n \mathbf{x}_n}^{+}}{\partial\mathcal{Q}_n }
    =-2(I-\mathcal{Q}_n\hat{\mathcal{C}}_{\mathbf{y}_n|\mathbf{x}_n}) \hat{\mathcal{C}}_{\mathbf{x}_n \mathbf{x}_n}^{-}
    \hat{\mathcal{C}}_{\mathbf{y}_n|\mathbf{x}_n}^{\mathrm{T}}
    +2\mathcal{Q}_n\mathcal{R}=0\\
    \Rightarrow 
&
     \mathcal{Q}_n = \hat{\mathcal{C}}_{\mathbf{x}_n \mathbf{x}_n}^{-}
    \mathcal{C}_{\mathbf{y}_n|\mathbf{x}_n}^{\mathrm{T}}
    \left(\hat{\mathcal{C}}_{\mathbf{y}_n|\mathbf{x}_n} \hat{\mathcal{C}}_{\mathbf{x}_n \mathbf{x}_n}^{-} \hat{\mathcal{C}}_{\mathbf{y}_n|\mathbf{x}_n}^{\mathrm{T}} + \mathcal{R} \right)^{-1},
\end{aligned}
\end{equation}
where $\hat{\mathcal{C}}_{\mathbf{x}_n \mathbf{x}_n}^{-}$ is the predictive kernel covariance operator and is calculated by \eqref{eq:eqAKKF_prediction_mean}, 
$\hat{\mathcal{C}}_{\mathbf{y}_n|\mathbf{x}_n}$ is 
the empirical likelihood operator and is calculated as
\begin{equation}
\label{eq:eqAKKF_update_C}
\begin{aligned}
\hat{\mathcal{C}}_{\mathbf{y}_n|\mathbf{x}_n} &
%= \hat{\mathcal{C}}_{\mathbf{y}_n \mathbf{x}_n}\hat{\mathcal{C}}^{-1}_{\mathbf{x}_n \mathbf{x}_n} 
= \Upsilon_n\left(
\Phi_n^{\mathrm{T}}\Phi_n + \lambda_K I \right)^{-1}\Phi_n^{\mathrm{T}}\\
&= \Upsilon_n\left(
\Gramx+ \lambda_K I \right)^{-1}\Phi_n^{\mathrm{T}}.
\end{aligned}
\end{equation}
Here, the Gram matrix $\Gramx=\Phi_n^{\mathrm{T}}\Phi_n$, and $\lambda_{{K}}$ is the regularization parameter to modify $\Gramx$. In this paper,  $\lambda_{{K}}$  is set to be 0. $\mathcal{R}$ is set as $\mathcal{R} = \kappa I$, $\kappa$ is used to approximate  the  covariance  of  the  residual  of  the  observation operator.

Combine \eqref{eq:eqAKKF_proposal_C} and \eqref{eq:eqAKKF_update_C}, we can have the following reductions,
\begin{equation}
\label{eq:eq73}
   \begin{aligned}
  & \hat{\mathcal{C}}_{\mathbf{x}_n \mathbf{x}_n}^{-} \hat{\mathcal{C}}_{\mathbf{y}_n|\mathbf{x}_n}^{\mathrm{T}}\\
   &= \left[  \Phi_n (\tilde{S}_{n-1}^{+} + V_n)  \Phi_n ^{\mathrm{T}} 
\right]  \left[\Upsilon_n \left(\Gramx +  \lambda_K  I\right)^{-1}
\Phi_n^{\mathrm{T}}\right] ^{\mathrm{T}}\\
&=   \Phi_n S_n^{-} 
  \left[ \left(\Gramx +  \lambda_K  I\right)^{-1}
\Phi_n^{\mathrm{T}} \Phi_n \right] ^{\mathrm{T}} \Upsilon_n^{\mathrm{T}}\\
%=& \Phi_n S_n^{-}  \Phi_n ^{\mathrm{T}}  \Phi_n \left[ \left(\Gramx +  \lambda_K  I\right)^{-1}\right] ^{\mathrm{T}}  \Upsilon_n^{\mathrm{T}},\\
& = \Phi_n S_n^{-} 
\left[ \left(\Gramx+  \lambda_K  I\right)^{-1}\Gramx \right] ^{\mathrm{T}}
\Upsilon_n^{\mathrm{T}}\\
&\overset{\lambda_K  =0}{=\joinrel=}\Phi_n S_n^{-} \Upsilon_n^{\mathrm{T}},
   \end{aligned} 
\end{equation}
\begin{equation}
\label{eq:eq74}
   \begin{aligned}
  &\mathcal{C}_{\mathbf{y}_n|\mathbf{x}_n} \mathcal{C}_{\mathbf{x}_n \mathbf{x}_n}^{-} \mathcal{C}_{\mathbf{y}_n|\mathbf{x}_n}^{\mathrm{T}}\\
  & =\left[\Upsilon_n \left(\Gramx + \lambda_K  I\right)^{-1}
\Phi_n^{\mathrm{T}}\right] \left(  \Phi_n S_n^{-}  \Phi_n ^{\mathrm{T}}
\right)  \left[\Upsilon_n \left(\Gramx +  \lambda_K  I\right)^{-1}
\Phi_n^{\mathrm{T}}\right] ^{\mathrm{T}}\\
&= \Upsilon_n \left[\left(\Gramx +  \lambda_K I\right)^{-1} \Gramx \right] S_n^{-} 
\left[ \left(\Gramx +  \lambda_K  I\right)^{-1}\Gramx  \right] ^{\mathrm{T}}
\Upsilon_n^{\mathrm{T}}\\
&\overset{\lambda_K  =0}{=\joinrel=}\Upsilon_n  S_n^{-} \Upsilon_n^{\mathrm{T}}.
   \end{aligned} 
\end{equation}
Substitute \eqref{eq:eq73} and \eqref{eq:eq74} into \eqref{eq:eqAKKF_update_Derivete}, the AKKF gain $\mathcal{Q}_n$ can be calculated as,
\begin{equation}
   \begin{aligned}
    \mathcal{Q}_n &= \hat{\mathcal{C}}_{\mathbf{x}_n \mathbf{x}_n}^{-}
    \mathcal{C}_{\mathbf{y}_n|\mathbf{x}_n}^{\mathrm{T}}
    \left(\hat{\mathcal{C}}_{\mathbf{y}_n|\mathbf{x}_n} \hat{\mathcal{C}}_{\mathbf{x}_n \mathbf{x}_n}^{-} \hat{\mathcal{C}}_{\mathbf{y}_n|\mathbf{x}_n}^{\mathrm{T}} + \mathcal{R} \right)^{-1}\\
    & = \Phi_n S_n^{-} \Upsilon_n^{\mathrm{T}} \left(\Upsilon_n  S_n^{-} \Upsilon_n^{\mathrm{T}} + \kappa I \right)^{-1}\\
    & = \Phi_n S_n^{-} \Upsilon_n^{\mathrm{T}} \left(\Upsilon_n^{\mathrm{T}}\right)^{-1}  \left[\Upsilon_n^{\mathrm{T}} \Upsilon_n  S_n^{-} \Upsilon_n^{\mathrm{T}}\left(\Upsilon_n^{\mathrm{T}}\right)^{-1}  + \Upsilon_n^{\mathrm{T}}\kappa I \left(\Upsilon_n^{\mathrm{T}}\right)^{-1}  \right]^{-1}\Upsilon_n^{\mathrm{T}}\\
    & = \Phi_n S_n^{-}  \left(\Gramy S_n^{-}  + \kappa I  \right)^{-1}\Upsilon_n^{\mathrm{T}}.
   \end{aligned} 
\end{equation}

%Substitution of \eqref{eq:eqAKKF_proposal_C} and \eqref{eq:eqAKKF_update_C} into \eqref{eq:eqAKKF_update_Derivete} then yields the $\mathcal{Q}_n$ is calculated as \eqref{eq:eqAKKF_update_Q}, the derivation details are presented in Appendix I.
%\begin{align}
%\label{eq:eqAKKF_update_Q}
%\mathcal{Q}_n =
%\Phi_n S_n^{-}  \left(G_{yy}  S_n^{-}  + \kappa I  \right)^{-1}\Upsilon_n^{\mathrm{T}}.
%\end{align}

The covariance operator $\hat{\mathcal{C}}_{\mathbf{x}_n \mathbf{x}_n}^{+}$ in \eqref{eq:eqAKKF_update_Cxx} is further derived as
\begin{equation}
   \begin{aligned}
   \hat{\mathcal{C}}_{\mathbf{x}_n \mathbf{x}_n}^{+}%=  {\rm{cov}}(\phi(x_n)-\hat{\mu}_{x_n}^{+}),\\
   %= & {\rm{cov}}\big\{ \phi(x_n)-\left[\hat{\mu}_{x_n}^{-} +\mathcal{Q}_n\left(\phi(y_n)- \hat{\mathcal{C}}_{y_n|x_n}\hat{\mu}_{x_n}^{-}-\mathcal{R}\right) \right]\big\},\\
   %&= {\rm{cov}}\left[\phi(x_n)-\hat{\mu}_{x,n}^{-}
   %-\mathcal{Q}_n\mathcal{C}_{y_n|x_n}\left(\phi(x_n)-\hat{\mu}_{x_n}^{-}\right)-\mathcal{Q}_n \mathcal{R}
   %\right]\\
   %= & {\rm{cov}}\left[\left(I -  \mathcal{Q}_n\hat{\mathcal{C}}_{y_n|x_n}\right)\left( \phi(x_n)-\hat{\mu}_{x_n}^{-}\right)-\mathcal{Q}_n \mathcal{R}\right],\\
   = & \left(I -  \mathcal{Q}_n\hat{\mathcal{C}}_{\mathbf{y}_n|\mathbf{x}_n}\right)
   \hat{\mathcal{C}}_{\mathbf{x}_n \mathbf{x}_n}^{-}
   \left(I -  \mathcal{Q}_n\hat{\mathcal{C}}_{\mathbf{y}_n|\mathbf{x}_n}\right)^{\mathrm{T}} + \mathcal{Q}_n \mathcal{R} \mathcal{Q}_n^{\mathrm{T}}\\
  = &   \hat{\mathcal{C}}_{\mathbf{x}_n \mathbf{x}_n}^{-}  - \hat{\mathcal{C}}_{\mathbf{x}_n \mathbf{x}_n}^{-}\hat{\mathcal{C}}_{\mathbf{y}_n|\mathbf{x}_n}^{\rm{T}} \mathcal{Q}_n^{\mathrm{T}}- \mathcal{Q}_n\hat{\mathcal{C}}_{\mathbf{y}_n|\mathbf{x}_n} \hat{\mathcal{C}}_{\mathbf{x}_n \mathbf{x}_n}^{-}\\
   &+\mathcal{Q}_n\hat{\mathcal{C}}_{\mathbf{y}_n|\mathbf{x}_n} \hat{\mathcal{C}}_{\mathbf{x}_n \mathbf{x}_n}^{-} \hat{\mathcal{C}}_{\mathbf{y}_n|\mathbf{x}_n}^{\rm{T}}\mathcal{Q}_n^{\mathrm{T}} + \mathcal{Q}_n \mathcal{R} \mathcal{Q}_n^{\mathrm{T}}\\
   = &   \hat{\mathcal{C}}_{\mathbf{x}_n \mathbf{x}_n}^{-}  -  \Phi_n S_n^{-} \Upsilon_n^{\mathrm{T}} \mathcal{Q}_n^{\mathrm{T}}- \mathcal{Q}_n \Upsilon_nS_n^{-}\Phi_n^{\mathrm{T}}+\mathcal{Q}_n\left(\Upsilon_n  S_n^{-} \Upsilon_n^{\mathrm{T}} + \mathcal{R}\right)\mathcal{Q}_n^{\mathrm{T}}\\
 = &   \hat{\mathcal{C}}_{\mathbf{x}_n \mathbf{x}_n}^{-}  -  \Phi_n S_n^{-} \Upsilon_n^{\mathrm{T}} \mathcal{Q}_n^{\mathrm{T}}- \mathcal{Q}_n \Upsilon_nS_n^{-}\Phi_n^{\mathrm{T}}\\
  &+\Phi_n S_n^{-} \Upsilon_n^{\mathrm{T}} \left(\Upsilon_n  S_n^{-} \Upsilon_n^{\mathrm{T}} + \kappa I \right)^{-1}\left(\Upsilon_n  S_n^{-} \Upsilon_n^{\mathrm{T}} + \kappa I \right)\mathcal{Q}_n\\
 = &   \hat{\mathcal{C}}_{\mathbf{x}_n \mathbf{x}_n}^{-} - \mathcal{Q}_n \Upsilon_n 
   S_n^{-}\Phi_n^{\mathrm{T}}.
\end{aligned} 
\end{equation}

%\end{appendices}

%\input{Preliminaries}

%\input{KBR in Dynamic System.tex}

%\input{KKPF}
%\input{Simulation}
%\input{Conclusion}

%\input{Reference}
%\clearpage

\nocite{*}
\bibliographystyle{IEEEtran}
\bibliography{main.bib}
\end{document}